\documentclass[12pt]{article}

\usepackage{amsthm,amsmath,amssymb}
\RequirePackage[authoryear]{natbib}
\RequirePackage[colorlinks,citecolor=blue,urlcolor=blue]{hyperref}

\usepackage{bm}
\usepackage{bbm}
\usepackage[noabbrev]{cleveref}
\usepackage{dsfont}
\usepackage{graphicx}
\usepackage{mathabx}
\usepackage{xfrac}
\usepackage{todonotes}
\usepackage{dsfont}
\usepackage{enumerate}
\usepackage{psfrag,epsf}

\newtheorem{theorem}{Theorem}
\newtheorem{lemma}{Lemma}
\newtheorem{proposition}{Proposition}
\newtheorem*{remark}{Remark}
\newcommand{\ba}{\matr a}
\newcommand{\barJ}{\widebar J}
\newcommand{\barL}{\widebar L}
\newcommand{\bB}{\matr B}
\newcommand{\bG}{\matr G}
\newcommand{\bI}{\matr I}
\newcommand{\bo}{\matr o}
\newcommand{\bO}{\matr O}
\newcommand{\bOmeg}{\matr \Omega}
\newcommand{\bpsi}{\boldsymbol \psi}
\newcommand{\bPsi}{\mathbf \Psi}
\newcommand{\bs}{\matr s}
\newcommand{\bSig}{\matr \Sigma}
\newcommand{\bx}{\matr x}
\newcommand{\bY}{\matr Y}
\newcommand{\bZ}{\matr Z}
\newcommand{\DOmeg}{\matr \Omega_{D}}
\newcommand{\htheta}{\hat{\theta}}
\newcommand{\kro}{\otimes}
\newcommand{\matr}[1]{\boldsymbol{#1}}
\newcommand{\mb}{\matr m}
\newcommand{\mE}{\mathbb E}
\newcommand{\mEt}{\widetilde{\mE}}
\newcommand{\pt}{\widetilde{p}}
\newcommand{\pta}{p_{\theta}}
\newcommand{\spa}{\,}
\newcommand{\ta}{\theta}
\newcommand{\tastar}{\theta^{\star}}
\newcommand{\Tr}{\operatorname{Tr}}
\newcommand{\tr}{^{\top}}
\newcommand{\Var}{\mathbb V}

\DeclareMathOperator{\Diag}{Diag} 
\DeclareMathOperator{\diag}{diag} 
\DeclareMathOperator*{\argmax}{arg\,max}
\DeclareMathOperator{\opvec}{vec} 

\DeclareRobustCommand{\bmrobust}[1]{\bm{#1}}
\newcommand{\blind}{0}

\begin{document}

\def\spacingset#1{\renewcommand{\baselinestretch}%
{#1}\small\normalsize} \spacingset{1}






\if0\blind
{
  \title{\bf Evaluating Parameter Uncertainty in the Poisson Lognormal Model with Corrected Variational Estimators}
  \author{Bastien~Batardière \hspace{.2cm}\\
    MIA Paris-Saclay, Université Paris-Saclay, AgroParisTech, INRAE,\\
    Julien Chiquet \hspace{.2cm} \\
    MIA Paris-Saclay, Université Paris-Saclay, AgroParisTech, INRAE\\
    and \\
    Mahendra~Mariadassou \\
    MaIAGE, Université Paris-Saclay, INRAE}
  \maketitle
} \fi

\bigskip

\begin{abstract} Count data analysis is essential across diverse fields, from
ecology and accident analysis to single-cell RNA sequencing (scRNA-seq) and
metagenomics. While log transformations are computationally efficient,
mo\-del-based approaches such as the Poisson-Log-Normal (PLN) model provide
robust statistical foundations and are more amenable to extensions. The PLN
model, with its latent Gaussian structure, not only captures overdispersion but
also enables correlation between variables and inclusion of covariates, making
it suitable for multivariate count data analysis. Variational approximations
are a golden standard to estimate parameters of complex latent variable models
such as PLN, maximizing a surrogate likelihood. However, variational estimators lack theoretical
statistical properties such as consistency and asymptotic normality. In this
paper, we investigate the consistency and variance estimation of PLN parameters
using M-estimation theory. We derive the Sandwich estimator, previously studied
in \cite{sandwich}, specifically for the PLN model. We compare this approach to
the variational Fisher Information method, demonstrating the Sandwich
estimator’s effectiveness in terms of coverage through simulation studies.
Finally, we validate our method on a scRNA-seq dataset.
\end{abstract}

\noindent%
\vfill

\newpage
\spacingset{1.75}

\section{Introduction}

Count data appear in many different fields, such as ecology, accident
analysis, single-cell ribonucleic acid sequencing (scRNA-seq), and metagenomics. For example,
researchers may be interested in estimating the impact of external covariates on the abundances of different species in a biome or on the expressions of different
genes in a cell.

Count data are hard to analyze as is, and transformations must be performed
beforehand in order to extract meaningful statistics. While log transformation
followed by Gaussian analyses is fast and widely used, it lacks robust
statistical foundations, and model-based approaches are much more suitable
\citep{NoLogTransform} to extensions and theoretical analyses. In particular, negative-binomial (NB) and Poisson-based
models are popular choices for modeling count data and have been extensively
used in RNAseq studies \citep[see \emph{e.g.}][]{DEseq2}. The NB distribution,
which uses Poisson emission law combined to a Gamma-distributed parameter to induce
overdispersion, is generally preferred to the standard Poisson distribution to
satisfy the overdispersion property (higher variance than mean value)
repeatedly observed in sequencing-based count data \citep[including scRNA-seq,
see][]{overdispersion}.

The (multivariate) Poisson-Log-Normal \citep[in short PLN,
see][]{Aitchison1989} model offers a general framework for multivariate count
data: it describes the dependencies
between counts by means of a latent Gaussian variable.
As a Poisson distribution with Log-Normal distributed parameters, PLN models
result in overdispersion, just like the NB. However, the underlying Gaussian
structure of the latent variable in the PLN model makes correlation between variables
a natural parameter of the model, unlike its NB counterpart. More generally, the PLN model falls in the
family of latent variable models (LVMs), and
more specifically of multivariate generalized linear mixed models
(mGLMMs), sometimes also called generalized linear latent variable models
\citep[GLLVM, ][]{gllvm}. In those models, the distribution of observed responses usually
belongs either to the exponential family (Bernoulli, Binomial,
Poisson, Negative-Binomial, with or without Zero-Inflation, etc.)
or to the exponential dispersion model, such as the Tweedie distribution
\citep{zhang2013tweedie}. Model parameters are
related to linear combinations
of latent variables and covariates through a simple
link function. Parameter estimation for common GLLVMs is efficiently
implemented in some packages \citep{gllvm,statsmodels}, making it a popular
option for multivariate count data. However, they usually rely on
the assumption of low-dimensional latent variables and, therefore, fail when the
latent variable has the same dimension as the data. For high-dimensional latent
variables, Monte Carlo methods are a potential, yet very
time-consuming, solution and exact likelihood methods become intractable. We rely instead
on surrogate likelihood approaches, more precisely, variational inference.

Variational approximations provide computational efficiency for a wide range of
latent variable models used in many fields \citep{blei2017variational} and are amenable to handling high-dimensional latent variables. Many studies have demonstrated the empirical
accuracy of the resulting estimates based on simulation studies
\citep[see, e.g.,][for models related to PLN and community
ecology]{Ormerod2012,hui2017variational,niku2019efficient}. Unfortunately, the general
theory about the statistical properties of variational estimates
is still very scarce and model-dependent. For some models, such as the stochastic block model \citep{HOLLAND1983109}, consistency and asymptotic normality \citep{asympSBM} can be derived for the variational estimators with mean-field approximation. However, such properties have not yet been established for the PLN model.
Besides, while consistency is crucial to ensure the accuracy of inference, asymptotic
normality is essential for deriving statistically valid confidence intervals, which
are fundamental for rigorous statistical interpretation and decision making.

When it is not possible to establish an asymptotic estimator of the variance of an estimator, one possibility is to use resampling-based generic methods, such as the Jackknife \citep{Jackknife,Miller}: the Jackknife is a non-parametric
estimate of variance (and bias), working under mild assumptions, and has
sometimes been used to estimate the variance of variational estimates \citep{bootstrapvar}. However, it is computationally intensive, as it requires running the inference method $n$ times, where $n$ is the
sample size. We rather follow the same track as in \citet{sandwich}, who use M-estimation theory
\citep{Huber1964Mestim,huber2011robust} to prove the asymptotic normality of a variational estimator and derive its asymptotic variance in the same manner as in \cite{vdV98}.
The corresponding variance estimator is known as the \textit{Sandwich}
estimator, named for its structure: it involves a product of three matrices, the two outermost of which are equal. We adopt this approach in this paper.

\paragraph*{Related works}

The approach of \citet{hall2011theory} is the closest to ours. They obtained
consistency and asymptotic normality for a Poisson mixed model that does not
account for correlation between counts and
considers only a single covariate. Although their approach could be readily adapted to handle multiple covariates,
the strong assumption of independence between counts is fundamental to their
analysis and cannot be relaxed.

Laplace Approximation (LA) has been widely used to infer the parameters of GLLVMs
\citep{LA1993Breslow,LA2004Huber,niku2019efficient}. Under some conditions on
the complete-log-likelihood, the resulting
estimator belongs to the class of M-estimators \citep{Huber1964Mestim} and can be proved to be asymptotically normal.
However, their framework only considers low-rank approximations of the
covariance matrix. Estimation of full-rank covariance matrices, such as the ones we consider, are
computationally intractable in the \verb|gllvm| package \citep{gllvm}.
Furthermore, even LA with low-rank approximations become computationally intensive when the
number of variables exceeds a thousand, and the number of observations reaches a
hundred, often requiring over an hour to process \citep{batardierepy}.

In \cite{stoehr2024composite}, the variational estimator is utilized as
the starting point of an EM algorithm, which maximizes a
composite
likelihood \citep{lindsay1988composite} of the PLN model. This approach yields an
estimator that is both consistent and asymptotically normal. However, the algorithm
requires estimating $\mathcal O(p^2)$ low-dimensional integrals at each iteration using Monte Carlo methods, making it computationally infeasible for datasets with more than a hundred variables.\\

This paper is organized as follows: in \Cref{sec:preliminaries}, we introduce
the PLN model and parameter inference based on the variational approximation of the likelihood.
\Cref{sec:varfishinf} presents a variational approximation of the Fisher Information and derives its formula. The Sandwich-based variance estimation is detailed in
\Cref{sec:sandwich-correction}. A comparison of both methods on synthetic data is conducted in
\Cref{sec:simu}, and the Sandwich-based variance estimates are applied to an
scRNA-seq dataset in \Cref{sec:application}.

\paragraph*{Notations}
Let $m$ and $p$ be positive integers. The vector space of all $m\times p$
matrices with real entries is denoted by $\mathcal M_{m,p}(\mathbb R)$. The subset of all
symmetric, positive, and definite $p \times p$-matrices over
$\mathbb{R}$ is denoted by $\mathcal{S}_{++}^p$ and $\matr I_p$ denotes the identity matrix of size $p$.
The trace operator $\Tr$ for a square matrix $A \in
\mathcal{M}_{p,p}(\mathbb{R})$ is defined as $\Tr(A) = \sum_{j=1}^p A_{jj}$.
$\Diag(x)$ denotes a diagonal matrix with the elements of vector $x$ on its
diagonal and is sometimes abbreviated to $\boldsymbol D_x$ in computations. Conversely, $\diag(A)$ represents a vector containing the diagonal
elements of the square matrix $A$. Let $\kro$ denotes the Kronecker operator, that is
    $$(A \otimes B)_{i j}=A_{i / / d, j / / q} B_{i \% d, j \% q},$$
for  integers $d$ and $q$, $A \in \mathcal M_{m,p}(\mathbb R), \, B \in \mathcal M_{d,q}(\mathbb R),\, i \in
\{1,\dots, dm\},\, j \in \{1,\dots, pq\}$, where $//$ and $\%$ denotes the
truncating integer division and remainder respectively. The matrix $C = A \odot B$ is obtained by multiplying the corresponding
elements of matrices $A$ and $B$ of the same size.
The vectorization operator for a matrix $A \in
\mathcal{M}_{m,p}(\mathbb{R})$ consists in stacking its column together and is defined as
$$
\operatorname{vec}(A)=\left[A_{1,1}, \ldots, A_{m, 1}, A_{1,2}, \ldots, A_{m, 2}, \ldots, A_{1, p}, \ldots, A_{m, p}\right]^{\mathrm{T}}.
$$
When applied to
matrices or vectors, simple functions like $\log, \exp$ or square
apply element-wise, and $1/A$ denotes the element-wise inverse of a matrix or vector. $\mathcal N_p(\boldsymbol \mu, \matr \Sigma)$ denotes a
Gaussian distribution with mean $\boldsymbol \mu \in \mathbb R^p$ and
covariance $\matr \Sigma \in \mathcal S_{++}^p$ and $\mathcal N_p(x;\boldsymbol
\mu, \matr \Sigma)$ denotes its density evaluated at point $x\in \mathbb R^p$.
For $x \in \mathbb R^p$, $B_\delta(x)$ denotes the ball with radius $\delta$
around $x$.

\section{Preliminaries}\label{sec:preliminaries}

\subsection{Model}
The multivariate Poisson lognormal model \citep[in short PLN,
see][]{Aitchison1989} relates some $p$-dimensional observation vectors
$\matr{Y}_i = (Y_{i1}, \dots, Y_{ip})\in\mathbb{N}^p$ to some  $p$-dimensional
vectors of Gaussian latent variables $\bZ_i\in\mathbb{R}^p$ with precision
matrix $\bOmeg \in \mathcal{S}_{++}^p$  (that is, covariance matrix $\bSig \triangleq \bOmeg^{-1}$). Our formalism of the PLN model resembles a
multivariate generalized linear model, where the main effect is due to a linear
combination of $m$ covariates $\bx_i\in\mathbb{R}^m$ (including an intercept), with the possibility to add an offset
$\boldsymbol{o}_i\in\mathbb{R}^p$ for the $p$ variables in each sample:
\begin{equation*}
    \begin{array}{rcl}
  \text{latent space } & \bZ_i \sim  \mathcal{N}_p(\boldsymbol{0}_p,\bOmeg^{-1}), & \\
  \text{observation space }  &   \bY_i | \bZ_i\sim\mathcal{P}\left(\exp\{\bo_i + \bx_i\tr\bB + \bZ_i\}\right) &  Y_{ij} \, | \, Z_{ij} \, \text{indep.}
  \end{array}
\end{equation*}
The $m\times p$ matrix $\bB$ is the latent matrix of regression parameters. The
latent covariance matrix $\bSig$ describes the underlying residual structure of
dependence between the $p$ variables. We denote by $\matr{Y}, \matr{O},
\matr{X}$ the matrices with respective sizes $n\times p, n\times p$ and
$n\times m$ stacking row-wise the vectors of counts, offsets, and covariates
(respectively $\bY_i, \bx_i$ and $\bo_i$). We also denote by $\matr{Z}$
the $n\times p$ matrix of unobserved latent Gaussian vectors $\bZ_i$. We
are interested in inferring $\theta = (\bOmeg, \bB)\in \boldsymbol{\Theta} = \mathbb S_p^{++}
\times \mathcal{M}_{p,m}(\mathbb R) \subset \mathbb R^{d}$ with $d = p(m+p)$.

Because the latent layer $\matr{Z}$ is not observed, the PLN model is an
incomplete data model in the sense of \cite{DEMP1977}, who introduced
the celebrated EM algorithm to perform maximum likelihood for such models. This
relies on the complete-data log-likelihood $\mathcal L(\theta; \matr{Y}, \matr{Z}) =
\log p_\theta(\matr{Y}, \matr{Z})$ which, in the case at hand, is given by
\begin{align*}
 \mathcal L(\theta; \matr{Y}, \matr{Z}) = & \Tr( \matr{Y}\tr [\bO + \matr{Z} + \matr{X}\bB]) -
\Tr(\matr{A}\tr \matr{1}_{n,p})  + K(\matr{Y}) \\
& + \frac{n}2\log|\matr{\Omega}| - \frac12 \Tr(\matr{Z} \matr{\Omega} \matr{Z}\tr) - \frac{np}{2} \log(2\pi)
\end{align*}
where $K(\matr{Y}) = -\sum_{i,j} \log(Y_{ij}!)$ and
$\matr{A} = (\matr{a}_1,\dots, \matr{a}_n)\tr = [A_{ij}]$ with
$\matr{a}_i = (A_{ij})_{1 \leq j \leq p}$ and
\begin{equation*}
 A_{ij} = e^{o_{ij} + \bx_i\tr\bB_j + Z_{ij}} = \mathbb{E}_\theta[Y_{ij} | Z_{ij}].
\end{equation*}

\subsection{Estimation by Variational Inference}
\label{sec:var-prox}

To maximize the marginal likelihood, the standard EM algorithm uses the following decomposition by integrating over the latent variables $\bZ$:
\begin{equation}
  \label{eq:loglik}
  \begin{aligned}
    \log p_\theta(\matr{Y}) \triangleq \mathcal L(\theta; \matr{Y}) & =  \log
    \frac{p_\theta(\matr{Z}, \matr{Y})}{p_\theta(\matr{Z} | \matr{Y})} =
    \int_{\matr{Z}} \log \frac{p_\theta(\matr{Z}, \matr{Y})}{p_\theta(\matr{Z}
    | \matr{Y})} p_\theta(\matr{Z} | \matr{Y}) d\matr{Z}  \\
    & = \mE_\theta \left(\mathcal L(\matr{\theta}; \matr{Y}, \matr{Z}) \right) - \mE_\theta \left(\log p_\theta(\matr{Z} | \matr{Y}) \right),
  \end{aligned}
\end{equation}
where $\mE_\theta$ is the expectation under the conditional distribution
$p_\theta(\matr{Z} | \matr{Y})$. However, for the PLN model, this decomposition is untractable
since the conditional distribution $p_\theta(\matr{Z}| \matr{Y})$ has no
closed-form. The use of surrogate variational approximation \citep[see,
e.g.][]{Jaa01} of this distribution is a popular and computationally efficient
choice, which consists in optimizing a lower bound of the log-likelihood, often referred to as the \textit{Evidence
Lower bound} (hereafter ELBO). To this end, we replace $p_\theta(\matr{Z}_i|
\matr{Y}_i)$ by a surrogate $\pt_{\bpsi_i}(\matr{Z}_i)$, picked among  Gaussian
distributions with diagonal covariance matrix (\emph{i.e.} independent
components) as follows:
\[
\pt_{\bpsi_i}(\bZ_i) = \mathcal N_p(\bZ_i; \mb_{i}, \diag(\bs_{i}^2)), \quad \pt_{ \bpsi}(\bZ) = \otimes_{i=1}^n \pt_{\bpsi_i}(\bZ_i).
\]
where $\bpsi = (\bpsi_1, \dots, \bpsi_n) \in \bPsi^n = \bPsi\times \dots \times
\bPsi$, and $\bpsi_i = (\mb_i, \bs_i) \in \bPsi$ with $\mb_i = (m_{ij})_{j=1
\dots p}$ and $\bs_i = (s_{ij})_{j=1\dots p}$ are vectors of $\mathbb R^{p}$. With a
slight abuse of notation, we identify $\pt_{\bpsi_i}$ with the
\textit{variational} parameters $\bpsi_i$ used to define it.

While various other choices could be made \citep[see,
e.g.,][]{blei2017variational}, we derive the ELBO by relying on the
Kullback-Lebler divergence between the true conditional distribution
$p_{\theta}(\matr{Z}_i|\matr{Y}_i)$  and its approximation
$\pt_{\bpsi_i}(\matr{Z}_i)$. It is defined as
\[
 KL(p_\theta(. | \matr{Y}) \| \pt_{\bpsi}(.) ) = \int_{\matr{Z}} \log \frac{p_\theta(\matr{Z} | \matr{Y})}{\pt_{\bpsi}(\matr{Z})} \pt_{\bpsi}(\matr{Z}) d\matr{Z}.
\]
Subtracting this (positive) quantity to the untractable Expression \eqref{eq:loglik} of the log-likelihood results after some rearrangements in an explicit ELBO:
\begin{align*}
 J(\bpsi, \theta)
 &= \mathcal L(\theta; \matr{Y}) - KL(p_\theta(\matr{Z}|\matr{Y}) || p_{\bpsi}(\matr{Z}))
\\
  & = \mEt_{\bpsi}\left(\mathcal L(\matr{\theta}; \matr{Y}, \matr{Z}) \right) - \mEt_{\bpsi} \left( \log p_\theta(\matr{Z} | \matr{Y}) \right) \leq \mathcal L(\theta; \matr{Y}),
\end{align*}
where $\mEt_{\bpsi}$ is the expectation under $\pt_{\bpsi}$. From this
expression, it is now obvious that the variational approach generalizes the EM
one by integrating over arbitrary (yet convenient) distribution.

In the case at hand, the ELBO -- also referred to as the
\textit{variational} log-likelihood -- is fully tractable. We denote
by $J_i(\theta, \bpsi_i)$ the term associated to the $i$-th
observation, with the following closed-form \citep[see e.g.][for a
detailed derivation]{ChiquetVersatile}:
\begin{multline}
  \label{eq:elbo_i}
 J_{i}(\theta, \bpsi_i) = \matr Y_i\tr [\bo_i + \mb_i + \bx_i\tr \bB] -
\tilde{\ba}_i\tr \matr 1_{p} + K(\matr Y_i) + \frac12 \log|\matr{\Omega}| \\ - \frac12
\mb_i\tr \matr{\Omega} \mb_i  - \frac12 \diag(\matr{\Omega})\tr \bs^2_i + \log |\diag(\bs_i)| + \frac{p}{2}
\end{multline}
where $\tilde{\matr{\! A}} =
(\tilde{\ba}_1,\dots,\tilde{\ba}_n)\tr =
[\tilde{A}_{ij}]$ is the variational counterpart to $\matr{A}$:
$$
\tilde{A}_{ij} = \mEt_{\bpsi_i} \left(e^{o_{ij} + \bx_i\tr \bB_{j} +
Z_{ij}}\right) = e^{o_{ij} + \bx_i\tr \bB_{j} + m_{ij} +
s_{ij}^2/2}.
$$
The ELBO corresponding to the full data set, obtained by summing the
$(J_i)_{i=1..n}$, has the following compact matrix form:
\begin{multline*}
 \barJ(\theta, \bpsi)
  = \Tr( \matr{Y}\tr [\matr{O} + \matr{M} + \matr{X}\bB]) -
\Tr(\tilde{\matr{A}}\tr \matr{1}_{n,p}) + K(\matr{Y}) + \frac{n}2\log|\matr{\Omega}|\\
- \frac12 \Tr(\matr{M} \matr{\Omega} \matr{M}\tr)   - \frac12 \Tr(\bar{\matr{S}}^2 \matr{\Omega}) +
\Tr(\log(\matr{S})\tr \matr{1}_{n,p}) + \frac12 np,
\end{multline*}
where $\matr{M} = [\mb_1, \dots, \mb_n]\tr$, $\matr{O} = [\bo_1, \dots, \bo_n]\tr$,
$\matr{S} = [\bs_1, \dots, \bs_n]\tr$,
$\bar{\matr{S}}^2 = \sum_{i=1}^n \diag(\bs_i^2)$.

The quantity $J(\theta,\bpsi)$ can be shown to be bi-concave in $\theta,\bpsi$
respectively. We can then derive a variational version of the EM, which
iterates until convergence between optimization of $J(\theta, \bpsi)$ with
respect to $\theta$ and then with respect to $\bpsi$, in order to obtain what
will be referred to as the \emph{variational estimator} of
$\theta$:
\begin{equation*}
  \left(\widehat{\theta}^{\text{ve}}, \widehat{\bpsi} \right) = \argmax_{\bpsi, \theta} \barJ(\bpsi, \theta).
\end{equation*}

We provide in \citet{ChiquetVersatile} and its accompanying
\verb|R/C++| package \textbf{PLNmodels} \citep{PLNmodels} an efficient
implementation of the variational EM algorithm for computing this
estimator. A Python package
\verb|pyPLNmodels|\footnote{\url{https://github.com/PLN-team/pyPLNmodels}}
\citep{batardierepy} is also available for large
scale problems, relying on \verb|Pytorch| \citep{pytorch}.

\section{Variational Fisher Information}\label{sec:varfishinf}

Generally speaking, there is no statistical guarantee for variational
estimators since we do not know how far the ELBO is from the log-likelihood. A
first naive approach would be to consider that the estimators obtained by
maximizing $J({\theta})$ have the same behavior as the maximum likelihood
estimator (MLE), defined by $\widehat{{\theta}}^{\text{mle}} =
\argmax_{{\theta}} \log \pta(\matr{Y})$. Desirable MLE properties are:
$i)$ unbiasedness; $ii)$ consistency; $iii)$ efficiency; and $iv)$ asymptotic
normality. For the latter, referring to $\theta^\star$ as the true model
parameter, one has
\[
\sqrt{n} (\widehat{{\theta}}^{\text{mle}} - \theta^\star) \xrightarrow{n \to \infty} \mathcal N(0, \mathcal I({\theta}^\star)^{-1}),
\]
where $\mathcal I$ is the Fisher information for a for a single sample, defined below, and closely related to the \textit{score function}:
\begin{equation}
  \label{sand:eq:fisher_info}
  \begin{aligned}
    \mathcal I({\theta}) & = \mE_{\theta}\left[ U_{\theta}(Y) U_{\theta}(Y)^\top  \right] = - \mE_{\theta} \left[ \frac{\partial^2 \mathcal L(\ta; Y)}{\partial
        {\theta}^2} \right],\\
     U_{\theta}(Y) & = \frac{\partial \mathcal L(\ta; Y)}{\partial {\theta}} \text{ (score function).} \\
  \end{aligned}
\end{equation}
In particular, asymptotic normality gives the approximation $\widehat{{\theta}}
\sim \mathcal N({\theta},  \widehat{\Var}[\widehat{{\theta}}])$ (where
$\widehat{\Var}[\theta]$ is any reasonable  estimate of
$\Var[\theta]$) which can be used to construct confidence intervals. Since $\mathcal I(\widehat{{\theta}})$ is a good approximation of
$\Var[\widehat{{\theta}}]^{-1}$ for MLE, a first naive straightforward approach
would be to derive the variational counterpart of the Fisher information matrix
$\mathcal I(\widehat{{\theta}}^{\text{ve}})$ to estimate
$\Var[\widehat{{\theta}}^{\text{ve}}]$ and use it to build variational confidence intervals.
We follow this path in this section.

Our naive baseline consists in directly computing the variational
counterpart of the Fisher information matrix~\eqref{sand:eq:fisher_info} and
plugging in the value of the variational estimate ${\theta}$ which maximizes
the ELBO to get $\widehat{\Var}[\widehat{{\theta}}^{\text{ve}}]$. This requires first
and second-order derivatives of variational log-likelihood with respect to the model parameters ${\theta} = (\matr{B},
\matr{\Omega})$ (gradient and Hessian matrix), the
computation of which is straightforward but tedious and thus postponed to
\Cref{sand:subsec:gradient}. We immediately derive a variational score and
variational Fisher information matrix from them, stated in \Cref{sand:prop:var-score} and \Cref{sand:prop:fisher-info}, respectively.

\begin{proposition}[Variational Score]
  \label{sand:prop:var-score}
  The variational approximation of the score is given by
  \begin{align*}
  \begin{split}
  \widetilde{U}({\theta}) \triangleq \frac{\partial
\barJ(\bpsi, {\theta})}{\partial {\theta}} = \opvec \left(\matr{X}^\top (\matr{Y}
- \matr{\widetilde{A}}), \frac{n}{2} \left[
\matr{\Omega}^{-1} - \frac{\matr{M}^\top\matr{M} +
\bar{\matr{S}}}{n} \right] \right).
  \end{split}
\end{align*}
\end{proposition}

\begin{proposition}[Variational Fisher Information]
  \label{sand:prop:fisher-info}
  The variational Fisher information matrix for a $n$-sample is given by
  \begin{align*}
  \widetilde{I}_n({\theta}) =
  \begin{pmatrix}
  (\matr{I}_p \kro
\matr{X}^\top)\Diag(\opvec(\matr{\widetilde{A}}))(\matr{I}_p \kro \matr{X}) & \matr{0} \\
  \matr{0} & \frac{n}{2}\matr{\Omega}^{-1} \kro
\matr{\Omega}^{-1}
  \end{pmatrix}.
  \end{align*}
\end{proposition}

\begin{remark}
 Let $\widetilde{A}_{.j}$ be the $j$-th column of $\matr{\widetilde{A}}$, then
 $(\matr{I}_p \kro \matr{X}^\top)\Diag(\opvec(\matr{\widetilde{A}}))(\matr{I}_p
 \kro \matr{X})$ is block-diagonal ($p$ square blocks of size $d$) with $j$-th
 block $\matr{X}^\top \Diag(\widetilde{A}_{.j}) \matr{X}$, which is
 positive definite as soon as $\matr{X}$ is full rank.
\end{remark}

A first estimator of the variance of $\widehat{{\theta}}^{\text{ve}}$, i.e,
$\widehat{\Var}[\widehat{{\theta}}^{\text{ve}}]$, is obtained by computing
$\widetilde I_n({\widehat{\theta}}^{\text{ve}})^{-1}$. In
particular, from the block-wise shape of $\widetilde I_n(\theta)^{-1}$, we get
\begin{equation}
  \label{sand:eq:variational-variance-1}
  \begin{aligned}
    \widehat{\Var}\left[\widehat{\matr{B}}_{kj}^{\text{ve}}\right] & = [(\matr{X}^\top \Diag(\widetilde{A}_{.j}) \matr{X})^{-1}]_{kk}\, , \\
    \widehat{\Var}\left[\widehat{\Omega}_{k\ell}^{\text{ve}}\right] & = \frac{2}{n}\widehat{\Omega}^{\text{ve}}_{kk}\widehat{\Omega}^{\text{ve}}_{\ell\ell}\, . \\
  \end{aligned}
\end{equation}

\begin{remark}
    Another commonly used method, not detailed in this paper, to estimate the variance in the context of missing data information and the EM algorithm is the Louis
approach \citep{louis1982finding}, which suggests a different estimator of
the Fisher matrix. The resulting variational Fisher information estimator for $\matr
B$, denoted $\widetilde I^{\text{louis}}_n({\matr B})$, satisfies
$\widetilde I^{\text{louis}}_n({\matr B}) \succeq \widetilde I_n(\matr B)$ with
the partial ordering over $\mathbb S_{++}$.
\end{remark}

\section{Sandwich Variance Based on M-estimation}
\label{sec:sandwich-correction}

In the previous section, we used the variational approximation as a
plug-in for Maximum Likelihood. We abusively relied on the property of
the MLE to build confidence intervals for variational estimators. In
the current section, we take a different path by following along the
lines of \citet{sandwich}, who interpret the variational estimator
as an M-estimator and use the classical theory of \citet{vdV98} to derive
confidence intervals.

\subsection{Profiled objective function}
Let $i \in \{1, \dots, n\}$. We recall that $J_i$ denotes the lower bound of the log-likelihood associated
with the $i$-th observation. It is obvious from its functional shape (Eq. (\ref{eq:elbo_i})) that $J_i$
is continuous in $(\theta,
\bpsi_i)$ for all $\matr{Y}_i$. Note $\widehat{\bpsi}_i = \argmax_{\psi}
J_{i}(\theta, \psi)$. We first state some properties relative to $\widehat{\bpsi}_i$.

\begin{proposition} \label{sand:prop:derivative-hatpsi}
 For all $\theta = (\matr{B}, \matr{\Omega}) \in
 \left(\mathcal{M}_{p,d}(\mathbb R)\times \mathbb S_p^{++}\right)$ and
 $\matr{Y}_i \in \mathbb N^p$, the map $\psi \mapsto J_{i}(\theta, \psi)$ is
 concave and thus admits a unique maximizer $\widehat{\matr \psi}_i$. Furthermore, the map
 $(\theta, \matr{Y}_i) \mapsto \widehat{\matr \psi}_i$ is $\mathcal{C}^\infty$ with first
 derivative w.r.t. $\theta$ given by
 \begin{equation} \label{sand:eq:gradinv}
   \nabla_\theta \widehat{\matr \psi}_i = - \left[ \nabla_{\matr \psi_i \matr
   \psi_i} J_{i}(\theta, \widehat{\matr \psi}_i) \right]^{-1} \nabla_{\matr
\psi_i \theta} J_{i}(\theta, \widehat{\matr \psi}_i).
\end{equation}
\end{proposition}

\begin{proof}
 The concavity claim is trivial. Regarding the smoothness of $(\theta,
 \matr{Y}_i) \mapsto \widehat{\matr \psi}_i$, note that $\widehat{\matr \psi}_i$ is implicitly defined
 by $\nabla_{\psi} J_{i}(\theta, \widehat{\matr \psi}_i) = 0$. The map $(\matr{Y}_i, \theta,
 \psi) \mapsto J_{i}(\theta, \psi)$ is $\mathcal{C}^{\infty}$. By the
 analytic implicit function theorem, $(\theta, \matr{Y}_i) \mapsto
 \widehat{\matr \psi}_i$ is thus also $\mathcal{C}^{\infty}$ and has first derivative
 with respect to $\theta$ given by \Cref{sand:eq:gradinv}.
\end{proof}

Define the profiled single data objective function as
\begin{equation*}
  L(\theta; \matr{Y}_i) \triangleq \sup_{\psi} J_{i}(\theta, \psi) = J_{i}(\theta, \widehat{\matr \psi}_i).
\end{equation*}
The variational estimate $\widehat{\theta}^{\text{ve}}$ is thus simply an M-estimator for the objective function
$$
\barL(\theta) = \frac{1}{n} \sum_{i=1}^n L(\theta; \matr{Y}_i).
$$
We hereafter drop the subscript $i$ to designate a generic observation
$Y$ and associated quantities, in particular, $J(\theta, \psi)$ corresponds to $J_i(\theta, \matr \psi_i)$.

\subsection{Consistency}

Let $M(\theta) = \mE^\star[L(\theta; Y)]$ be the expectation of $L(\theta; Y)$ under the true parameter $\theta^\star$.

\paragraph{Assumptions}

\begin{itemize}
 \item[(A1)] Assume that the parameter space $\boldsymbol{\Theta}$ of $\theta$
     is compact. This is the case if $i)$ the eigenvalues of $\matr{\Omega}$
     are bounded away from $0$ and $\infty$ and $ii)$ the coordinates of
     $\matr{B}$ are either bounded or if we extend $\mathbb R$ to $[-\infty,
     +\infty]$ and embed it with the metric $d(x, y) = |\arctan(x) -
     \arctan(y)|$.
 \item[(A2)] Assume that the variational parameter space $\boldsymbol{\Psi}$ of the variational parameters $\psi$ is bounded.
\end{itemize}

The following proposition ensures that $J$ is smooth enough for
$\htheta^{\text{ve}}$ to inherit standard consistency properties of
M-estimators.

\begin{proposition} \label{sand:prop:smoothness-J}
 Under assumptions $(A1)$ and $(A2)$ the following properties hold:
 \begin{itemize}
 \item[(i)] $\theta \mapsto J(\theta, \psi)$ is continuous for almost every $Y$;
 \item[(ii)] $\widehat{\theta}^{\text{ve}}$ is in a compact set with probability tending to one;
 \item[(iii)] the function $J(\theta, \psi)$ has a measurable and integrable (w.r.t. Y) local envelope function:
\begin{equation*}
\forall \delta >0, \spa \forall \theta'\in \mathbb R^d, \spa
\sup_{\psi \in \boldsymbol{\Psi}}\sup_{\theta \in \mathcal{B}_\delta(\theta')} J(\theta, \psi) \text{ is } p_{\tastar}\text{-integrable}.
\end{equation*}
 \end{itemize}
\end{proposition}

\begin{proof}
 Properties $(i)$ and $(ii)$ are trivial: $J(\theta, \psi)$ is
 $\mathcal{C}^{\infty}$ (and thus continuous) and $\widehat{\theta}^{\text{ve}}$
 belongs to the compact parameter space $\boldsymbol{\Theta}$. Concerning
 $(iii)$, note that $J(\theta, \psi)$ decomposes as $J(\theta, \psi) =
 J^{(1)}(\theta, \psi) + J^{(2)}(\theta, \psi)$ where
 \begin{align*}
     J^{(1)}(\theta, \psi) & = Y^\top [\matr{o} + \matr{m} + \matr{B}\tr\matr{x}] - \sum_{j=1}^p\log(Y_{j}!) \\
  J^{(2)}(\theta, \psi) & = -\matr{\widetilde{a}}^\top \matr{1}_{p} + \frac12 \log|\matr{\Omega}| - \frac12 \matr{m}^\top \matr{\Omega} \matr{m} - \frac12 \diag(\matr{\Omega})^\top \matr{s}^2 + \frac 1 2\log (\matr{s^2})^\top \matr{1}_p - \frac{p}{2}.
 \end{align*}
 For all $\theta' \in \boldsymbol{\Theta}$ and $\delta > 0$, by boundedness of $\mathcal{B}_\delta(\theta') $ and $\boldsymbol{\Psi}$ there exists constants
 $C^{(1)}_{\matr{x}}$ and $C^{(2)}$ such that:
 \[
    \sup_{\stackrel{\theta \in \mathcal{B}_\delta(\theta')}{\psi \in \boldsymbol{\Psi}}} J^{(1)}(\theta, \psi) \leq C^{(1)}_{\matr{x}} \|Y\|_1 - \sum_{j=1}^{p}\log(Y_{j}!)  \quad \text{and} \quad \sup_{\stackrel{\theta \in \mathcal{B}_\delta(\theta')}{\psi \in \boldsymbol{\Psi}}} J^{(2)}(\theta, \psi) \leq C^{(2)}.
 \]
 Now we show that $Y \mapsto C^{(1)}_{\matr{x}} \|Y\|_1 - \sum_{j=1}^p\log(Y_{j}!)) + C^{(2)}$ is
 $p_{\theta^\star}$-integrable. The
 first two moments of $Y$ are finite from \citet{chiquet2018variational}, so
 that $\left\| Y\right\|_1$ has finite expectation. Since $|\sum_{j=1}^p\log(Y_{j}!))| \leq
 \sum_{j=1}^p Y_j^2$, we have that $-\sum_{j=1}^p\log(Y_{j}!)$ is $p_{\theta^{\star}}$-integrable.
\end{proof}

\begin{theorem}[Consistency of $\htheta^{\text{ve}}$]
\label{sand:thm:Consistency}
Under assumptions $(A1)-(A2)$, assume that the map $M(\theta)$ attains a finite
global maximum at $\bar{\theta}$ (which can be different from $\theta^\star$).
Then $\widehat{\theta}^{\textnormal{ve}} \to \bar{\theta}$ under $p_{\theta^\star}$.
\end{theorem}

\begin{proof}
This is a direct application of Theorem 1 from
\citet{sandwich}, which is itself a direct
consequence of Theorem 5.14 from \citet{vdV98}, as
items (i)-(iii) from
Prop.~\ref{sand:prop:smoothness-J} correspond to
their regularity conditions (A1)-(A3).
\end{proof}

\begin{remark}
 The value $M(\theta)$ is intractable, making it challenging to ascertain whether the global
maximum $\bar{\theta}$ is unique and coincides with $\theta^\star$.
However, simulations detailed in \Cref{subsubsec:simu_rmse} suggest that $\widehat{\theta}$ is unbiased so that
 $\bar{\theta} = \theta^\star$.
\end{remark}

\begin{remark}
 Boundedness of $\boldsymbol{\Psi}$ is sufficient but may not be necessary. Further
 analyses are required to check whether the local envelope has finite expectation
 when $\boldsymbol{\Psi}$ is unbounded.
\end{remark}

\subsection{Asymptotic normality}

We first state a few additional properties satisfied by $ \widehat{\matr \psi}
$ and $J$. Those are smoothness properties (Lipschitz conditions) that ensure
that $\htheta^{\text{ve}}$ inherits standard convergence properties of
M-estimators.

\begin{proposition} \label{sand:prop:normality-assumptions}
Under assumptions (A1)-(A2), the following propositions holds
\begin{enumerate}
 \item There exist $\kappa(Y) \in L_1(p_{\theta^\star})$ and $\delta > 0$ such that $\forall \theta \in \mathcal{B}_\delta(\bar{\theta})$, $p_{\theta^\star}$ a.s.:
 \[
 \left| (\nabla^2_{\theta\theta} J - (\nabla^2_{\theta\psi} J)(\nabla^2_{\psi\psi} J)^{-1}(\nabla^2_{\psi\theta} J)^\top)(\theta, \widehat{\matr \psi}) \right| \leq \kappa(Y)
 \]
 \item There exist $r > 0,s(Y)>0, b_1(Y)$ and $b_2(Y)$ such that
 \begin{itemize}
  \item[(i)] For all $Y \in \mathbb N^p$ and $\theta \in \mathcal{B}_r(\bar{\theta})$, $\widehat{\matr \psi}(\theta, Y) \in \mathcal{B}_{s(Y)}(\widehat{\matr \psi}(\bar{\theta}, Y))$;
  \item[(ii)] For all $Y \in \mathbb N^p$ and $\theta_1, \theta_2 \in \mathcal{B}_r(\bar{\theta})$ and $\psi_1, \psi_2 \in \mathcal{B}_{s(Y)}(\widehat{\matr \psi}(\bar{\theta}, Y))$,
  \[
   |J(\theta_1, \psi_1) - J(\theta_2, \psi_2)| \leq b_1(Y) (\|\theta_1 - \theta_2\| + \|\psi_1 - \psi_2\|);
  \]
  \item[(iii)] For all $\theta_1, \theta_2 \in \mathcal{B}_r(\bar{\theta})$, $\|\widehat{\matr \psi}(\theta_1, Y) - \widehat{\matr \psi}(\theta_2, Y)\| \leq b_2(Y) \| \theta_1 - \theta_2 \|$
  \item[(iv)] $b_1$ and $b_1 b_2 \in L_2(p_{\theta^\star})$
 \end{itemize}
\end{enumerate}
\end{proposition}

\begin{proof}
Note that the expression $(\nabla^2_{\theta\theta} J - (\nabla^2_{\theta\psi}
J)(\nabla^2_{\psi\psi} J)^{-1}(\nabla^2_{\psi\theta} J)^\top)(\theta, \psi)$
has no terms depending on $Y$ and is continous in $(\theta, \psi)$. It is thus
uniformly bounded on the compact $\mathcal{B}_\delta(\bar{\theta}) \times
\boldsymbol{\Psi}$ by a constant not depending on $Y$, which proves $1$.
Concerning 2, (i) and (iii) are direct consequences of
Propositon~\ref{sand:prop:derivative-hatpsi}. Close examination of $\left[
\nabla_{\psi \psi} J(\theta, \psi) \right]^{-1} \nabla_{\psi \theta} J(\theta,
\psi)$ reveals no term depending on $Y$ and is thus uniformly bounded by some
constant $C$ on $\mathbb N^p \times \mathcal{B}_r(\bar{\theta}) \times
\boldsymbol{\Psi}$. This means in turn that the quantity $\nabla_\theta
\widehat{\matr \psi}(\theta, Y)$ is itself bounded by $C$, and we can thus set $b_2(Y) = C$ and
$s(Y) = Cr$ to prove (i) and (iii). For (ii), we rely on the same decomposition
of $J$ into $J^{(1)}$ and $J^{(2)}$ used in the proof of
Prop.~\ref{sand:prop:smoothness-J}. $J^{(2)}$ does not depend on $Y$ and we
thus have $|J^{(2)}(\theta_1, \psi_1) - J^{(2)}(\theta_2, \psi_2)| \leq D
(\|\theta_1 - \theta_2\| + \|\psi_1 - \psi_2\|)$ for $D =
\sup_{\mathcal{B}_r(\bar{\theta}) \times \boldsymbol{\Psi}}
\max(\|\nabla_\theta J^{(2)}(\theta,\psi) \|_{\infty}, \| \nabla_\psi J^{(2)}(\theta,\psi) \|_{\infty}) <
\infty$. Note that $D$ is finite by smoothness of $J^{(2)}$ on the compact
$\mathcal{B}_r(\bar{\theta}) \times \boldsymbol{\Psi}$ and does not depend on
$Y$. For $J^{(1)}$, note that:
\[
J^{(1)}(\theta_1, \psi_1) - J^{(1)}(\theta_2, \psi_2) = Y^\top(\matr{m}_1 - \matr{m}_2) + Y\tr (\matr{B}_1 - \matr{B}_2)\tr \matr{x}
\]
We can thus set $b_1(Y) = D + \max(\|Y\|_\infty, \|Y\|_\infty\|\matr
x\|_\infty)$ to satisfy (ii). Finally note that $\|Y\|_{\infty} \leq \sum_{j=1}^p|Y_j| \in
L_2(p_{\theta^\star})$, which proves (iv).
\end{proof}
\begin{theorem}[Normality of $\htheta^{\text{ve}}$] \label{sand:thm:normality}
Under assumptions (A1)-(A2), suppose $\widehat{\theta}^{\textnormal{ve}} \xrightarrow[]{p_{\theta^\star}} \bar{\theta}$ where $\bar{\theta}$ is an argmax of $M(\theta)$. Then,
\[
\sqrt{n}(\widehat{\theta}^{\text{ve}} - \bar{\theta})  \xrightarrow[]{d} \mathcal{N}(0, V(\bar{\theta}))
\]
where $V(\theta) = C(\theta)^{-1} D(\theta) C(\theta)^{-1}$ for
\begin{align*}
 C(\theta) & = \mE[\nabla_{\theta\theta} L(\theta; Y) ] \\
 D(\theta) & = \mE\left[(\nabla_{\theta} L(\theta; Y)) (\nabla_{\theta} L(\theta; Y))^\top \right] \\
\end{align*}
\end{theorem}

\begin{proof}
This is a direct application of Theorem 2 from
\citet{sandwich}, as
Prop.~\ref{sand:prop:derivative-hatpsi} ensures
their regularity conditions (B1)-(B2)
and Prop.~\ref{sand:prop:normality-assumptions}
correspond to their regularity
conditions (B3)-(B4). This result is a direct
consequence of Theorem
5.23 from \citet{vdV98}.
\end{proof}

\subsection{Effective computations}

Remember that $M(\theta)$ is not available in closed form so that $C(\theta)$
and $D(\theta)$ from \Cref{sand:thm:normality} are not available
explicitely. We can however approximate them from $J$ using the chain rule as
\[
 \nabla_{\theta\theta} L(\theta; Y) = \left[ \nabla_{\theta\theta} J - \nabla_{\theta\psi} J (\nabla_{\psi\psi} J)^{-1} \nabla_{\theta\psi} J^\top \right] (\theta, \widehat{\matr \psi}(\theta, Y))
\]
and we thus compute $C(\theta)$ and $D(\theta)$ using the estimators $\widehat{C}_n(\htheta)$ and $\widehat{D}_n(\htheta)$ where
\begin{align*}
 \widehat{C}_n(\theta) & = \frac{1}{n} \sum_{i=1}^n \left[
 \nabla_{\theta\theta} J_{i} - \nabla_{\theta\matr \psi_i} J_{i} (\nabla_{\matr
\psi_i\matr \psi_i} J_{i})^{-1} \nabla_{\theta\matr \psi_i} J_{i}^\top
\right](\theta, \widehat{\matr \psi}_i), \\
 \widehat{D}_n(\theta) & = \frac{1}{n} \sum_{i=1}^n \left[ \nabla_{\theta} J_{i} \nabla_{\theta} J_{i}^\top \right](\theta, \widehat{\matr \psi}_i).
\end{align*}
Let us now expand each of those matrices in turn. We will only focus on the regression parameters $\boldsymbol{B}$. A similar formula can be derived for $\matr{\Omega}$.

\begin{remark}
    Unlike the variational Fisher information, the Sandwich variance matrix $V(\theta)$ is not
    block-wise. This implies that strictly speaking, the entire matrix
    $V(\theta)$ and consequently $C(\theta)^{-1}$ must be computed even if the sole objective is
    to determine the variance of the regression coefficient estimator. However,
    inverting the matrix $C(\theta)$ is impractical in practice because it
    is of size $mp^3\times mp^3$. This process becomes computationally
    prohibitive when $p>15$ and $m>1$.
    We choose to approximate the Sandwich variance matrix as block-wise (and thus neglect the cross terms between $\matr B$ and $\matr \Omega$) and
    compute only the term $V(\matr B)$, which is computationally feasible as
    it only requires inverting the matrix $C(\matr B) \in \mathcal
    M_{mp,mp}(\mathbb R)$. Simulations detailed in
    \Cref{subsubsec:simu_coverage} suggest that this approximation does not
    deteriorate the variance estimation.
\end{remark}

\paragraph{Formula for $\widehat{C}_n(\boldsymbol{B})$ and $\widehat{D}_n(\boldsymbol{B})$}
In the following, we denote $\matr{D}_{\widetilde{\!\matr{a}}_i}$ (resp
$\matr{D}_{\matr{s}_i}$) the diagonal matrix of size $p$ with diagonal
$\widetilde{\matr{a}}_i$ (resp. $\matr{s}_i$), and $\DOmeg =
\bI_p \odot \bOmeg$.
\begin{proposition}[Expression of $\widehat{D}_n(\boldsymbol{B})$] \label{sand:thm:Dhat}
After vectorizing over $\boldsymbol{B}$, one has
\[
 \widehat{D}_n(\boldsymbol{B}) = \frac{1}{n} \sum_{i=1}^n \left[ (\matr{Y}_i -
 \!\widetilde{\!\matr{a}}_i)(\matr{Y}_i - \!\widetilde{\!\matr{a}}_i)^\top
\right] \kro \left(\matr{x}_i \matr{x}_i^\top\right) \in \mathbb R^{mp \times
mp}.
\]
\end{proposition}

\begin{proof}
    This form is a direct consequence of the derivative given in \Cref{sand:prop:inverse-deriv-1-Theta} in \Cref{sand:app:Chat_s}.
\end{proof}

\begin{remark}
 Each of the $n$ small matrices $\left[ (\matr{Y}_i -
\matr{\widetilde{a}}_i)(\matr{Y}_i - \matr{\widetilde{a}}_i)^\top \right] \kro
\left(\matr{x}_i \matr{x}_i^\top \right)$ is at most of rank $1$.
$\widehat{D}_n(\boldsymbol{B})$ is therefore at most of rank $n$. \end{remark}

\begin{proposition}[Expression of $\widehat{C}_n$] \label{sand:thm:Chat}
    With the previous notations and after vectorizing over $\boldsymbol{B}$,
\[
    \widehat{C}_n(\boldsymbol{B}) = - \frac{1}{n} \sum_{i=1}^n \left(\widehat{\matr{\Sigma}}^{\text{ve}} +
  \matr{D}_{\widetilde{\matr{a}}_i}^{-1} + \matr{D}_{\matr{s}_i}^4 \left(
\matr{I}_p + \matr{D}_{\matr{s}_i}^2 (\matr{D}_{\widetilde{\matr{a}}_i} +
\DOmeg)\right)^{-1} \right)^{-1} \kro \left( \matr{x}_i
\matr{x}_i^\top  \right).
\]
The computational complexity of determining such a matrix of size $mp\times mp$
is $\mathcal O(np^3)$.
\end{proposition}

\begin{proof}
    Using \Cref{sand:prop:inverse-deriv-2-theta-theta} and averaging over $i = \{1,\dots,n\}$ gives the result.
\end{proof}

\begin{remark}
    The overall complexity of computing $V(\boldsymbol{B})$ is dominated by either the computation of $C(\boldsymbol{B})$ (if $n \geq m^3$) or by its inversion and is thus $\mathcal{O}(\max(m^3, n) p^3)$.
\end{remark}

\section{Simulation Study}\label{sec:simu}
In this section, we assess, on synthetic data, the consistency of $\widehat{\theta}^{\text{ve}}$
and the asymptotic normality of the regression coefficient $\widehat{\matr
B}^{\text{ve}}$. The variance estimates for asymptotic normality include the
Sandwich estimate, as presented in \Cref{sand:thm:normality}, and the
Variational Fisher Information, as detailed in
\Cref{sand:eq:variational-variance-1}.

\paragraph{Data generation} Let $n \in \{1000,2000,3000\}, p\in \{50,100,200\}$ and $m \in \{1,2,3,4\}$.
For each pair $(m,p)$, we simulate $K = 100$ design matrices $\boldsymbol X_{(m,p,k)}$ and datasets $\boldsymbol Y_{(m,p,k)}$ with
parameters $\theta^{\star}_{(m,p)}$ for $k = 1,\dots, K$.

The $n \times m$ design matrix $\matr X_{(m,p,k)}$
comprises independent rows $\bx_i$ following a multinomial
distribution such that $\mathbb P(x_{ir} =1) = 1/m$ for $r \in \{1,\dots, m\}$.
The parameter $\matr B^{\star}_{(m,p)}$ is sampled as
\begin{equation*}
    \matr B^{\star}_{(m,p)_{kj}} \overset{\text{\footnotesize indep}}{\sim} \mathcal N(2,1), \quad k \in \{1,\dots,m\}, \spa j \in \{1, \dots, p\},
\end{equation*}
ensuring that $\matr X_{(m,p,k)} \matr B^{\star}_{(m,p)}$ exhibits independent
Gaussian entries centered on $2$ with unit variance.
The covariance matrix $\matr \Sigma_{(m,p)}^{\star}$ is defined as the sum of a random Toeplitz matrix and the identity matrix:
\begin{equation*}
    \matr \Sigma_{(m,p)_{kj}}^{\star} = \mathds{1}_{j=k} + \rho_{(m,p)}^{|j-k|}, \quad \rho_{(m,p)} \sim \mathcal U([0.8,0.95]), \spa (j,k) \in \{1,\dots,p\}^2.
\end{equation*}
The random parameter $\rho$ controls the amount of correlation between variables. Offsets ($\matr
O$) are not considered in these simulations and are set to a zero matrix of
dimension $n \times p$.

\subsection{Numerical experiments}
\subsubsection{Consistency}\label{subsubsec:simu_rmse}
\begin{figure}[htb!]
        \begin{center}
            \includegraphics[width=\linewidth]{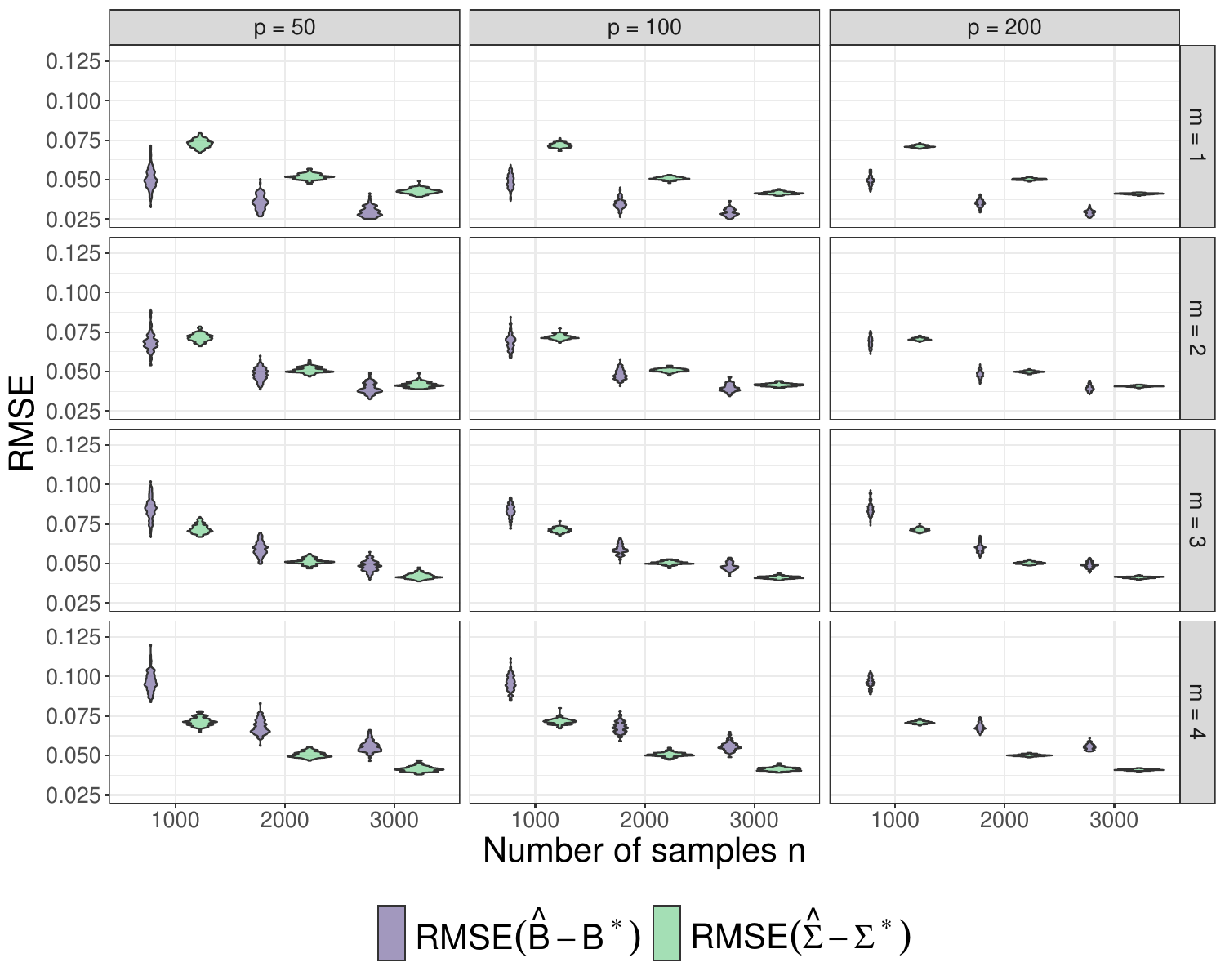}
            \caption{Root mean squared error as a function of the number of
covariates $m$, the number of variables $p$ and the number of samples $n$.
Each violin plot is based on $K=100$ synthetic datasets.}
            \label{fig:rmse}
        \end{center}
\end{figure}

\begin{figure}[htb!]
        \begin{center}
            \includegraphics[width=\linewidth]{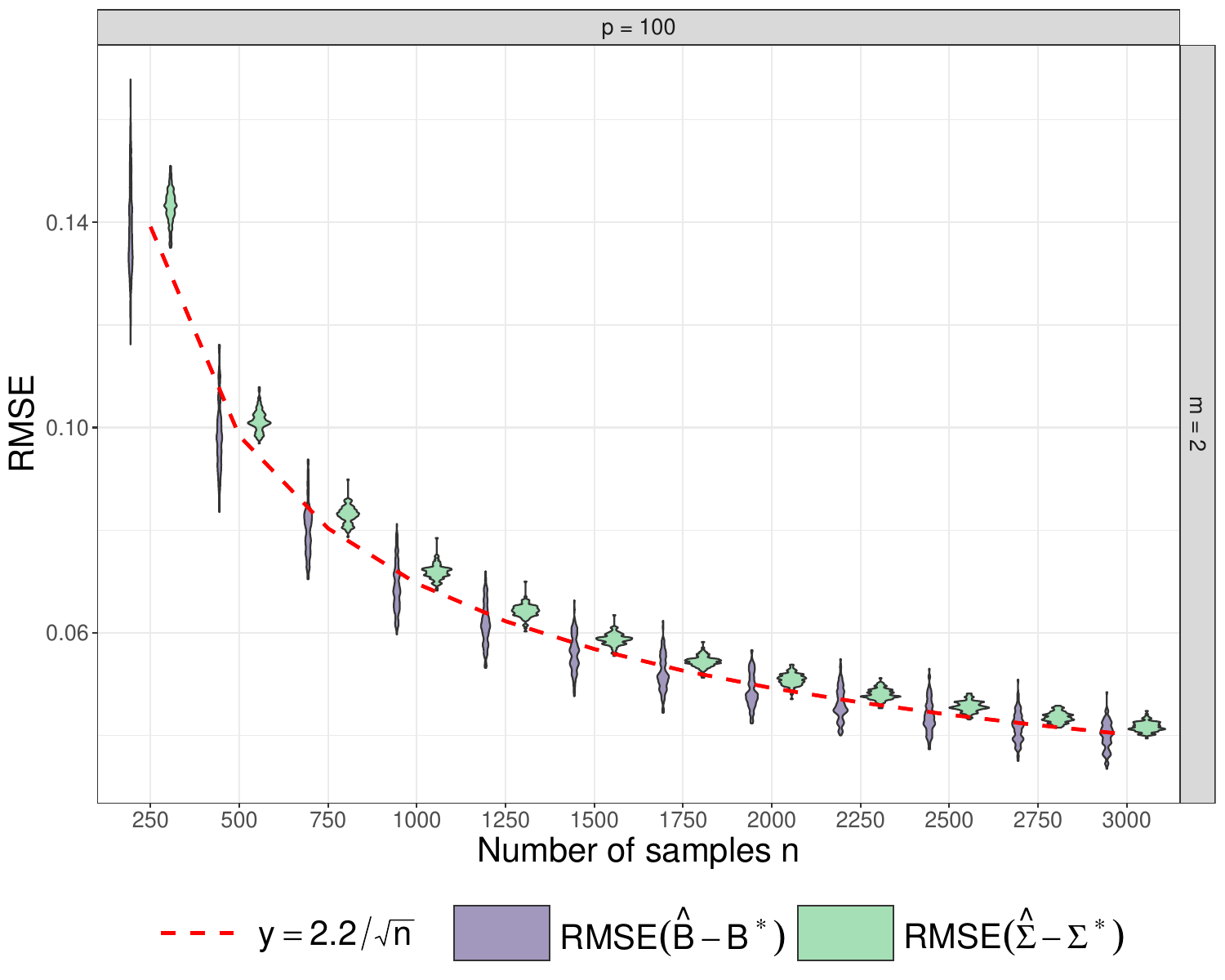}
            \caption{Root mean squared error  as a function of the number of
                samples $n$ for $p=100$ and $m=2$. The red dashed line
                corresponds to the equation $y=2.2/\sqrt{n}$. Each violin plot is based on $K=100$ synthetic datasets.}
            \label{fig:rmse_loglog}
        \end{center}
\end{figure}

To evaluate the consistency, we compute the Root Mean Squared Error (RMSE)
between the variational estimate $(\widehat{\matr B}, \widehat{\matr \Sigma})$ and
true parameters $(\matr B^{\star}, \matr \Sigma^{\star})$ and show the results in \Cref{fig:rmse} and \Cref{fig:rmse_loglog}.
\Cref{fig:rmse} suggests that the variational estimates have either small or no bias, as the RMSE keeps decreasing towards small values with increasing $n$. They are also robust to $p$ and $m$ as the RMSE increases with neither $p$ nor $m$ in the considered range. To futher investigate the convergence rate, we fixed $p=100$ and $m=2$ and considered a finer grid over $n$. The results, shown in \Cref{fig:rmse_loglog} show that the observed convergence rate is $\mathcal{O}(n^{- \sfrac{1}{2}})$, which is expected for unbiased estimators.

\subsubsection{Asymptotic normality of the regression coefficients}

We examine the standardized estimates:
\begin{equation}
    \left(\widehat{B}_{kj} - B^{\star}_{kj}\right)\bigg /\sqrt{\widehat{\mathbb V}(\widehat{B}_{kj})} \label{eq:standard}
\end{equation}
where $\widehat{B}_{kj}$ and $B_{kj}^{\star}$ denote the $(k,j)$ entries of the
estimated regression parameter $\widehat{\bB}$ and the
true regression parameter $\bB^{\star}$, respectively.
$\widehat{\mathbb{V}}(\widehat{B}_{kj})$ represents the estimated variance of
the associated parameter, computed using either the variational Fisher
information or the Sandwich-based variance. The standardized estimates are expected to be approximately $\mathcal{N}(0, 1)$.

\begin{figure}
        \begin{center}
            \includegraphics[width=\linewidth]{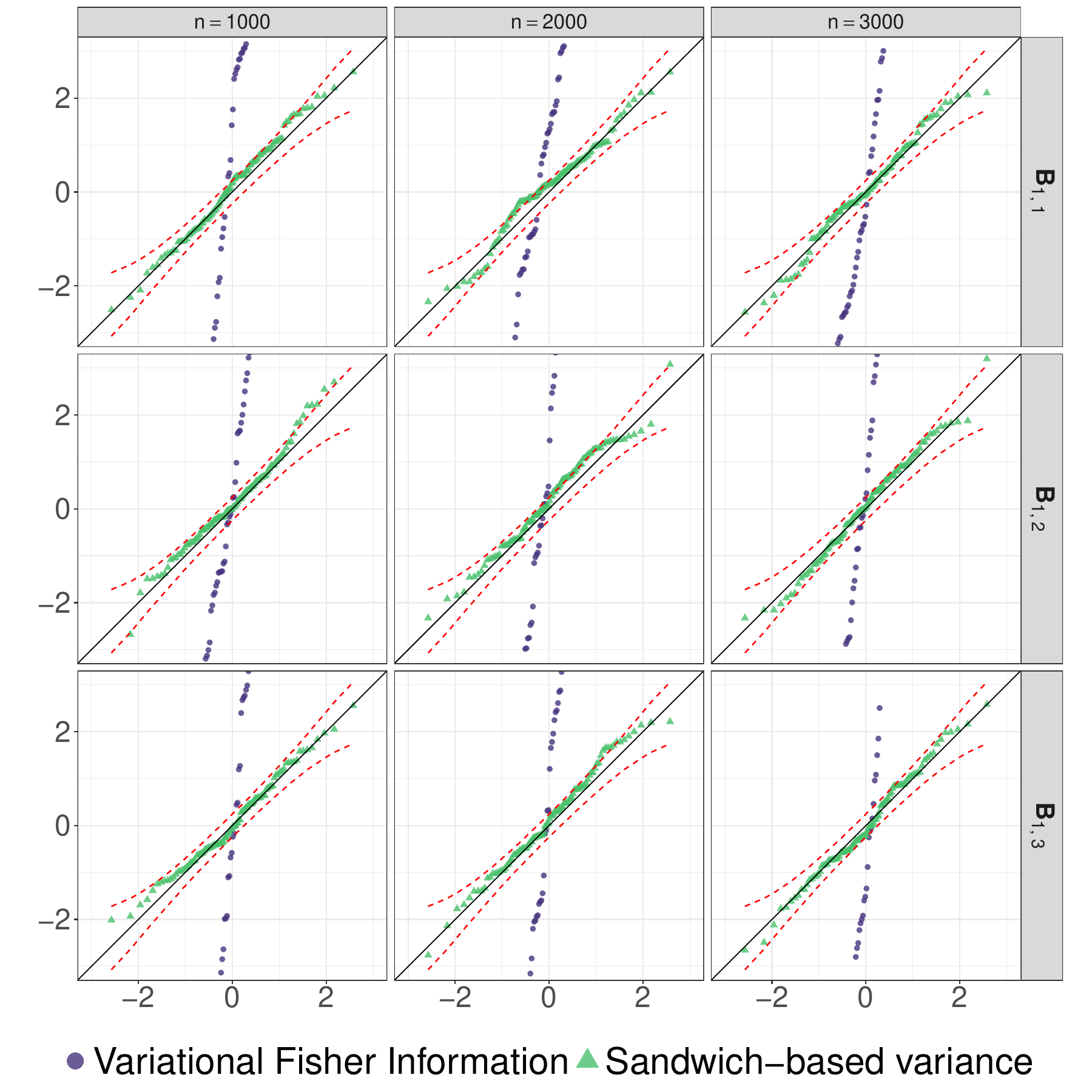}
            \caption{
QQ plots for three standardized regression coefficients
based on $K = 100$ simulations with $p = 200$ and $m
= 3$ on synthetic data.
The QQ plots are computed using standardized coefficients with variance
estimated with Variational Fisher Information (circle purple points) and
Sandwich-based variance (triangle green points). The black straight line
corresponds to the $y = x$ line, and the red dashed lines show $95\%$
intervals for QQ plots of 100 centered Gaussian points with unit variance.
For clarity, the plots are restricted to the range -3 to 3.
}\label{fig:qqplots}
        \end{center}
\end{figure}

\begin{figure}
        \begin{center}
            \includegraphics[width=\linewidth]{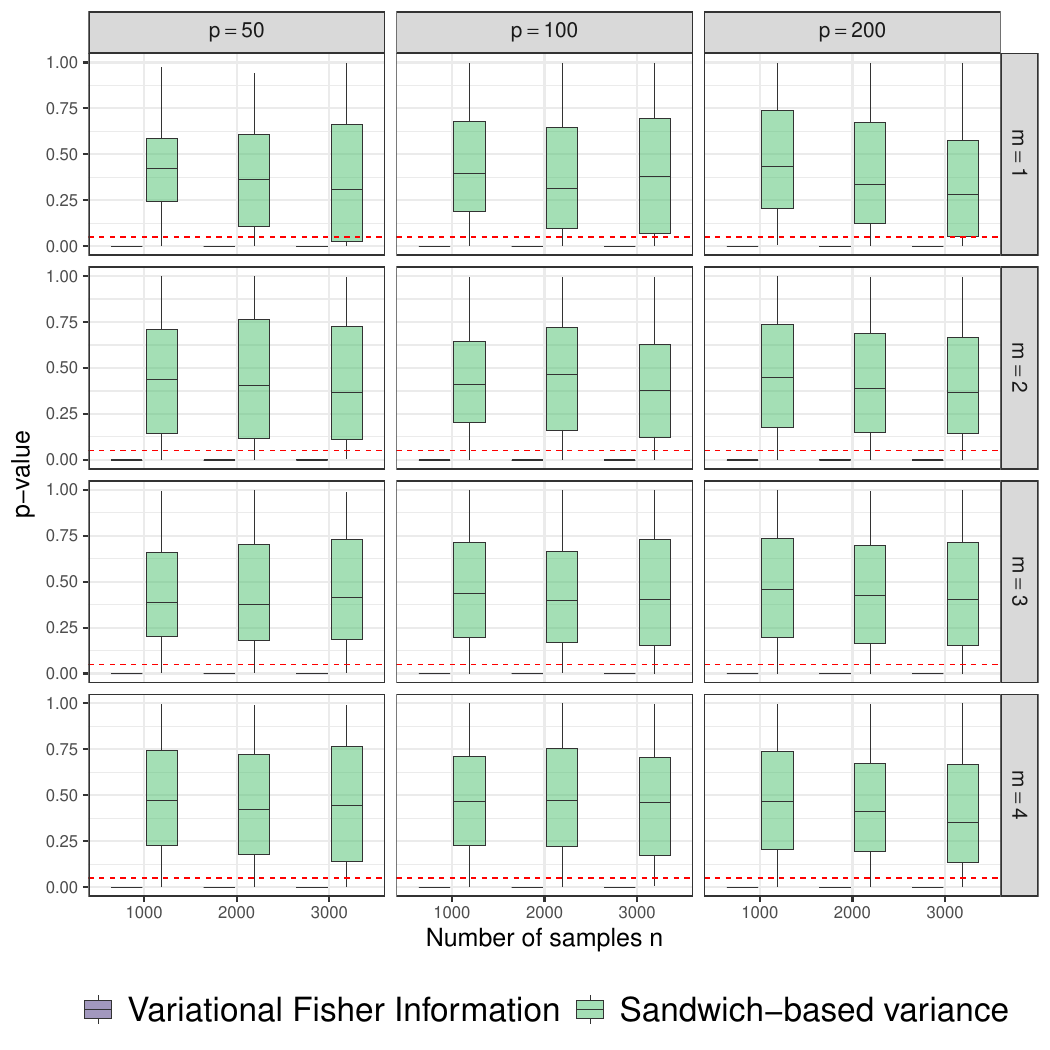}
            \caption{
Boxplots of $p$-values from the one-sample Kolmogorov-Smirnov test of equality $\mathcal{N}(0, 1)$ distribution applied to the standardized estimates
over $K = 100$ simulations on synthetic data. Each boxplot is constructed from the $m \times p$
$p$-values. The $p$-values are computed using two methods: standardizing estimates with the
variational Fisher Information (purple boxplots) and Sandwich-based variance (green points).
The dotted red line represents the $\alpha = 5\%$ significance threshold.
}
            \label{fig:ks}
        \end{center}
\end{figure}

A qualitative assessment of the distribution of the standardized
estimates across simulations ($k=1,\dots,K$) is shown in
\Cref{fig:qqplots}. A quantitative assessment is provided in
\Cref{fig:ks}, which includes boxplots of $p$-values from a one-sample Kolmogorov-Smirnov (KS) test of equality to a $\mathcal{N}(0, 1)$ distribution.

The QQ plots in \Cref{fig:qqplots} demonstrate that the variational Fisher
Information consistently underestimates the variance,
whereas the Sandwich-based method provides accurate variance estimates.
This is evidenced by the alignment of the standardized estimates' quantiles
with the theoretical quantiles.

The quantitative KS $p$-value plots in \Cref{fig:ks} further illustrate the
non-normality of the standardized estimates when using the variational Fisher
Information for variance estimation. In contrast, the Sandwich-based variance
estimation results in high $p$-values in the KS test, indicating compatibility
with the standard Gaussian.

\subsubsection{Coverage}\label{subsubsec:simu_coverage}

\begin{figure}
        \begin{center}
            \includegraphics[width=\linewidth]{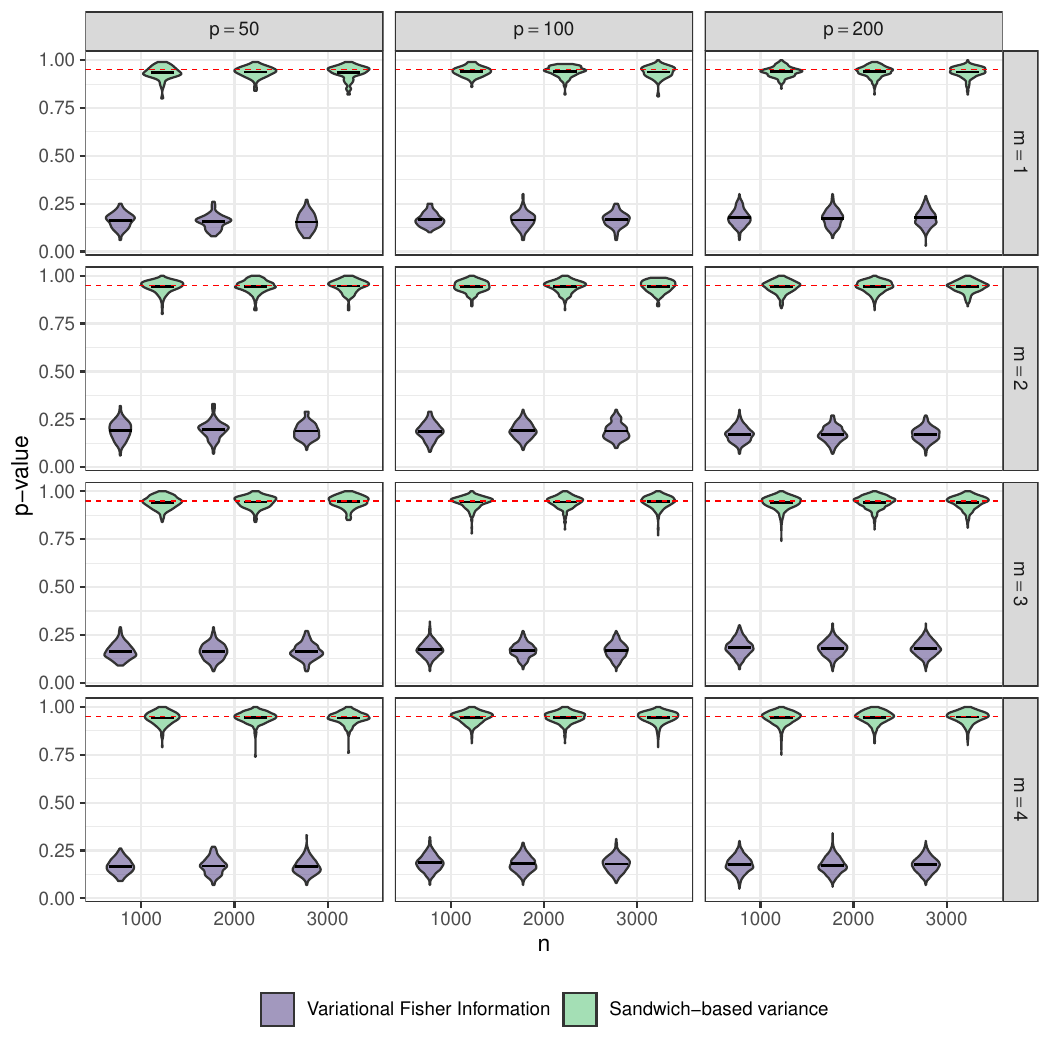}
            \caption{ 95 \% coverage for $K=100$ datasets for the parameter
                $\matr B$, as a function of the number of covariates $m$, the
                number of samples $n$, and the number of variables $p$.
            The dotted red line represents the $95\%$ threshold, and horizontal
        black lines correspond to the mean of the $p$-values. }
            \label{fig:coverage}
        \end{center}
\end{figure}

To assess the coverage of confidence intervals build using the sandwich estimator of the variance. We compute the proportion of standardized estimate entries from
\Cref{eq:standard} that lie within the $95\%$ confidence interval $\left[-\phi_{0.975},
\phi_{0.975}\right]$, where
$\phi_{0.975}$ represents the 97.5\% quantile of the Gaussian
distribution. The results are presented in \Cref{fig:coverage}.

In all scenarios, the Variational Fisher Information consistently underestimates the variance
associated with the variational estimator, resulting in confidence intervals that are too narrow and have poor coverage. By contrast, the sandwich-based estimation accurately assesses the variance and provides confidence intervals with a nominal coverage of 95\%,
even when the number of estimated parameters $m \times p$ is large ($m = 4, p
= 200$) and approaches the number of samples ($n = 1000$).

\section{Application to the scMARK dataset}\label{sec:application}
In this section, we derive confidence intervals for the regression coefficient of the PLN model on the scMARK dataset.
The latter is a benchmark for scRNA-seq data
designed to serve as an scRNA-seq equivalent of the MNIST dataset: each cell
is labeled by one of the 28 possible cell types.
The dataset consists of $n=19 998$ samples (cells) and $p = 14 059$ features (gene
expression).

Due to memory limitations, we cannot address the problem with the full set of $p =
14059$ features. We fit our model instead on a dataset reduced to $n = 5000$ samples and $p = 300$ features. The $5000$ samples are randomly
selected among the $m=3$ most prevalent cell types, and the $300$ genes are
those with the largest variance in the original dataset. We assign a unique
number to each cell type and denote $K_i$ as the number representing the cell
type of cell $i$ for $i \in \{1,\dots,n\}$.
These cell types are used as covariates, such that for each cell $i$,
$$x_{ik} = \mathds{1}_{k = K_i}.$$
The resulting regression coefficient $\matr B$ is a matrix of size $m \times p$,
where each column $\matr B_k$ corresponds to the mean of the latent variable
associated with cell type $k$.

\begin{figure}
        \begin{center}
            \includegraphics[width=\linewidth]{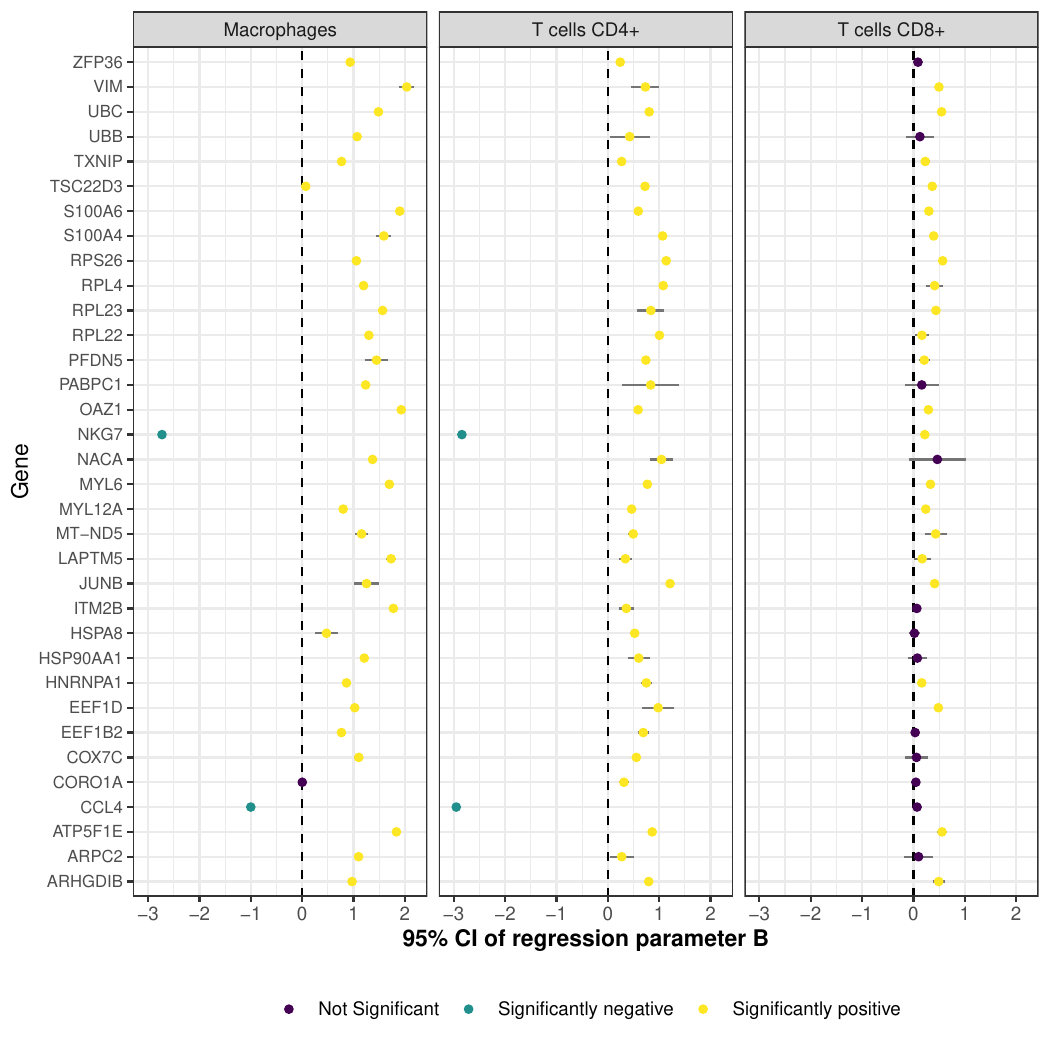}
            \caption{
            Forest plot of confidence intervals obtained with Sandwich-based standardization for regression coefficients of $m = 3$ cell groups (T CD8+,
            T CD4+, and Macrophage cells) from the scMARK dataset, reduced to $n = 5000$ samples (cells) and $p
            = 300$ variables (genes).
The y-axis denotes the gene names.
For visualization purposes, we displayed only the regression variables with
estimated coefficients that are positive and less than $0.6$ for the mean of T
cells CD8+. This resulted in $34$ coefficients out of the $m \times p = 900$
available.
}
            \label{fig:cov_real}
        \end{center}
\end{figure}

We display a forest plot of estimated coefficients along with their confidence
intervals in \Cref{fig:cov_real}.
The forest plot highlights genes that are differentially expressed
in CD8+ T cells but not in macrophages or CD4+ T cells. Notably,
the gene NKG7 is differentially expressed in CD8+ T cells, which is
consistent with the findings reported in \cite{li2022nkg7}.

\paragraph*{Complexity}
The complexity of fitting the PLN model using variational approximation is
$\mathcal{O}(n(p(m + p)))$. Nevertheless, the \verb|pyPLNmodels| package is optimized
 to handle high-dimensional data through various optimization techniques efficiently.
Consequently, fitting the PLN model on the scMARK dataset with $n=5000$, $m=3$,
and $p=300$ takes up to $8$ seconds on a GPU (RTX A5000) and $48$ seconds on a CPU.
A detailed analysis is performed in \cite{batardierepy}.
Regarding the Sandwich estimation, with a complexity of
$\mathcal{O}(\max(n,m^3)p^3)$, it took $22$ seconds to compute on both GPU and CPU.

\section*{Fundings}
  Bastien Bartardière and Julien Chiquet are supported by the French
  ANR grant ANR-18-CE45-0023 Statistics and Machine Learning for
  Single Cell Genomics (SingleStatOmics).

\section*{Acknowledgments}

We thank Julien Stoehr for his valuable suggestions on the graphs representing
$p$-values and QQ plots, as well as for the fruitful discussions regarding the PLN
model.

\bibliographystyle{plainnat}
\bibliography{biblio}

\begin{thebibliography}{34}
\providecommand{\natexlab}[1]{#1}
\providecommand{\url}[1]{\texttt{#1}}
\expandafter\ifx\csname urlstyle\endcsname\relax
  \providecommand{\doi}[1]{doi: #1}\else
  \providecommand{\doi}{doi: \begingroup \urlstyle{rm}\Url}\fi

\bibitem[Aitchison and Ho(1989)]{Aitchison1989}
John Aitchison and CH~Ho.
\newblock The multivariate poisson-log normal distribution.
\newblock \emph{Biometrika}, 76\penalty0 (4):\penalty0 643--653, 1989.

\bibitem[Batardiere et~al.(2024)Batardiere, Kwon, and Chiquet]{batardierepy}
Bastien Batardiere, Joon Kwon, and Julien Chiquet.
\newblock {pyPLNmodels: A Python package to analyze multivariate
  high-dimensional count data}.
\newblock October 2024.

\bibitem[Bickel et~al.(2013)Bickel, Choi, Chang, and Zhang]{asympSBM}
Peter Bickel, David Choi, Xiangyu Chang, and Hai Zhang.
\newblock {Asymptotic normality of maximum likelihood and its variational
  approximation for stochastic blockmodels}.
\newblock \emph{The Annals of Statistics}, 41\penalty0 (4):\penalty0 1922 --
  1943, 2013.

\bibitem[Blei et~al.(2017)Blei, Kucukelbir, and McAuliffe]{blei2017variational}
David~M Blei, Alp Kucukelbir, and Jon~D McAuliffe.
\newblock Variational inference: A review for statisticians.
\newblock \emph{Journal of the American statistical Association}, 112\penalty0
  (518):\penalty0 859--877, 2017.

\bibitem[Breslow and Clayton(1993)]{LA1993Breslow}
N.~E. Breslow and D.~G. Clayton.
\newblock Approximate inference in generalized linear mixed models.
\newblock \emph{Journal of the American Statistical Association}, 88\penalty0
  (421):\penalty0 9--25, 1993.

\bibitem[Chen et~al.(2018)Chen, Wang, and Erosheva]{bootstrapvar}
Yen-Chi Chen, Y.~Samuel Wang, and Elena~A. Erosheva.
\newblock On the use of bootstrap with variational inference: theory,
  interpretation, and a two-sample test example.
\newblock \emph{The Annals of Applied Statistics}, 12\penalty0 (2):\penalty0
  846--876, 2018.

\bibitem[Chiquet et~al.(2018)Chiquet, Mariadassou, and
  Robin]{chiquet2018variational}
J~Chiquet, M~Mariadassou, and S~Robin.
\newblock Variational inference for probabilistic poisson pca.
\newblock \emph{The Annals of Applied Statistics}, 2018.

\bibitem[Chiquet et~al.(2021)Chiquet, Mariadassou, and Robin]{ChiquetVersatile}
J~Chiquet, M~Mariadassou, and S~Robin.
\newblock The poisson-lognormal model as a versatile framework for the joint
  analysis of species abundances.
\newblock \emph{Frontiers in Ecology and Evolution}, 9, 2021.

\bibitem[Chiquet et~al.(2024)Chiquet, Mariadassou, Robin, and
  Gindraud]{PLNmodels}
Julien Chiquet, Mahendra Mariadassou, Stéphane Robin, and Fran\c{c}ois
  Gindraud.
\newblock \emph{PLNmodels: Poisson Lognormal Models}, 2024.
\newblock R package version 1.2.0.

\bibitem[Choudhary and Satija(2022)]{overdispersion}
Saket Choudhary and Rahul Satija.
\newblock Comparison and evaluation of statistical error models for scrna-seq.
\newblock \emph{Genome biology}, 23\penalty0 (1):\penalty0 27, January 2022.
\newblock \doi{10.1186/s13059-021-02584-9}.

\bibitem[Dempster et~al.(1977)Dempster, Laird, and Rubin]{DEMP1977}
A.~P. Dempster, N.~M. Laird, and D.~B. Rubin.
\newblock Maximum likelihood from incomplete data via the {EM} algorithm.
\newblock \emph{Journal of the Royal Statistical Society: Series B},
  39:\penalty0 1--38, 1977.

\bibitem[Efron and Stein(1981)]{Jackknife}
B.~Efron and C.~Stein.
\newblock {The Jackknife Estimate of Variance}.
\newblock \emph{The Annals of Statistics}, 9\penalty0 (3):\penalty0 586 -- 596,
  1981.

\bibitem[Hall et~al.(2011)Hall, Ormerod, and Wand]{hall2011theory}
Peter Hall, John~T Ormerod, and Matt~P Wand.
\newblock Theory of gaussian variational approximation for a poisson mixed
  model.
\newblock \emph{Statistica Sinica}, pages 369--389, 2011.

\bibitem[Holland et~al.(1983)Holland, Laskey, and Leinhardt]{HOLLAND1983109}
Paul~W. Holland, Kathryn~Blackmond Laskey, and Samuel Leinhardt.
\newblock Stochastic blockmodels: First steps.
\newblock \emph{Social Networks}, 5\penalty0 (2):\penalty0 109--137, 1983.

\bibitem[Huber(1964)]{Huber1964Mestim}
Peter~J. Huber.
\newblock {Robust Estimation of a Location Parameter}.
\newblock \emph{The Annals of Mathematical Statistics}, 35\penalty0
  (1):\penalty0 73 -- 101, 1964.

\bibitem[Huber and Ronchetti(2011)]{huber2011robust}
Peter~J Huber and Elvezio~M Ronchetti.
\newblock \emph{Robust statistics}.
\newblock John Wiley \& Sons, 2011.

\bibitem[Huber et~al.(2004)Huber, Ronchetti, and Victoria-Feser]{LA2004Huber}
Philippe Huber, Elvezio Ronchetti, and Maria-Pia Victoria-Feser.
\newblock {Estimation of Generalized Linear Latent Variable Models}.
\newblock \emph{Journal of the Royal Statistical Society Series B: Statistical
  Methodology}, 66\penalty0 (4):\penalty0 893--908, 2004.

\bibitem[Hui et~al.(2017)Hui, Warton, Ormerod, Haapaniemi, and
  Taskinen]{hui2017variational}
Francis~KC Hui, David~I Warton, John~T Ormerod, Viivi Haapaniemi, and Sara
  Taskinen.
\newblock Variational approximations for generalized linear latent variable
  models.
\newblock \emph{Journal of Computational and Graphical Statistics}, 26\penalty0
  (1):\penalty0 35--43, 2017.

\bibitem[Jaakkola and Jordan(1997)]{Jaa01}
Tommi~S. Jaakkola and Michael~I. Jordan.
\newblock A variational approach to {B}ayesian logistic regression models and
  their extensions.
\newblock In \emph{Proceedings of the Sixth International Workshop on
  Artificial Intelligence and Statistics}, volume~R1, pages 283--294, 1997.

\bibitem[Li et~al.(2022)Li, Corvino, Nowlan, Aguilera, Ng, Braun, Cillo, Bald,
  Smyth, and Engwerda]{li2022nkg7}
Xian-Yang Li, Dillon Corvino, Bianca Nowlan, Amelia~Roman Aguilera, Susanna~S
  Ng, Matthias Braun, Anthony~R Cillo, Tobias Bald, Mark~J Smyth, and
  Christian~R Engwerda.
\newblock Nkg7 is required for optimal antitumor t-cell immunity.
\newblock \emph{Cancer Immunology Research}, 10\penalty0 (2):\penalty0
  154--161, 2022.

\bibitem[Lindsay(1988)]{lindsay1988composite}
Bruce~G Lindsay.
\newblock Composite likelihood methods.
\newblock \emph{Comtemporary Mathematics}, 80\penalty0 (1):\penalty0 221--239,
  1988.

\bibitem[Louis(1982)]{louis1982finding}
Thomas~A Louis.
\newblock Finding the observed information matrix when using the em algorithm.
\newblock \emph{Journal of the Royal Statistical Society: Series B
  (Methodological)}, 44\penalty0 (2):\penalty0 226--233, 1982.

\bibitem[Love et~al.(2014)Love, Huber, and Anders]{DEseq2}
Michael~I Love, Wolfgang Huber, and Simon Anders.
\newblock Moderated estimation of fold change and dispersion for rna-seq data
  with deseq2.
\newblock \emph{Genome biology}, 15\penalty0 (12):\penalty0 1--21, 2014.

\bibitem[Miller(1974)]{Miller}
Rupert~G. Miller.
\newblock The jackknife--a review.
\newblock \emph{Biometrika}, 61\penalty0 (1):\penalty0 1--15, 1974.

\bibitem[Niku et~al.(2019{\natexlab{a}})Niku, Brooks, Herliansyah, Hui,
  Taskinen, and Warton]{niku2019efficient}
Jenni Niku, Wesley Brooks, Riki Herliansyah, Francis~KC Hui, Sara Taskinen, and
  David~I Warton.
\newblock Efficient estimation of generalized linear latent variable models.
\newblock \emph{PLOS ONE}, 14\penalty0 (5), 2019{\natexlab{a}}.

\bibitem[Niku et~al.(2019{\natexlab{b}})Niku, Hui, Taskinen, and Warton]{gllvm}
Jenni Niku, Francis~KC Hui, Sara Taskinen, and David~I Warton.
\newblock gllvm: Fast analysis of multivariate abundance data with generalized
  linear latent variable models in r.
\newblock \emph{Methods in Ecology and Evolution}, 10\penalty0 (12):\penalty0
  2173--2182, 2019{\natexlab{b}}.

\bibitem[O'Hara and Kotze(2010)]{NoLogTransform}
Robert O'Hara and D.J. Kotze.
\newblock Do not log-transform count data.
\newblock \emph{Methods in Ecology and Evolution}, 1:\penalty0 118--122, 2010.

\bibitem[Ormerod and Wand(2012)]{Ormerod2012}
J.~T. Ormerod and M.~P. Wand.
\newblock Gaussian variational approximate inference for generalized linear
  mixed models.
\newblock \emph{Journal of Computational and Graphical Statistics}, 21\penalty0
  (1):\penalty0 2--17, 2012.

\bibitem[Paszke et~al.(2019)Paszke, Gross, Massa, Lerer, Bradbury, Chanan,
  Killeen, Lin, Gimelshein, Antiga, Desmaison, Kopf, Yang, DeVito, Raison,
  Tejani, Chilamkurthy, Steiner, Fang, Bai, and Chintala]{pytorch}
Adam Paszke, Sam Gross, Francisco Massa, Adam Lerer, James Bradbury, Gregory
  Chanan, Trevor Killeen, Zeming Lin, Natalia Gimelshein, Luca Antiga, Alban
  Desmaison, Andreas Kopf, Edward Yang, Zachary DeVito, Martin Raison, Alykhan
  Tejani, Sasank Chilamkurthy, Benoit Steiner, Lu~Fang, Junjie Bai, and Soumith
  Chintala.
\newblock Pytorch: An imperative style, high-performance deep learning library.
\newblock In \emph{Advances in Neural Information Processing Systems 32}, pages
  8024--8035. 2019.

\bibitem[Seabold and Perktold(2010)]{statsmodels}
Skipper Seabold and Josef Perktold.
\newblock statsmodels: Econometric and statistical modeling with python.
\newblock In \emph{9th Python in Science Conference}, 2010.

\bibitem[Stoehr and Robin(2024)]{stoehr2024composite}
Julien Stoehr and Stephane~S. Robin.
\newblock Composite likelihood inference for the poisson log-normal model,
  2024.

\bibitem[{van der Vaart}(1998)]{vdV98}
A.~{van der Vaart}.
\newblock \emph{Asymptotic statistics}, volume~27 of \emph{Cambridge Series in
  Statistical and Probabilistic Mathematics}.
\newblock 1998.

\bibitem[Westling and McCormick(2019)]{sandwich}
T.~Westling and T.~H. McCormick.
\newblock Beyond prediction: A framework for inference with variational
  approximations in mixture models.
\newblock \emph{Journal of Computational and Graphical Statistics}, 28\penalty0
  (4):\penalty0 778--789, 2019.

\bibitem[Zhang(2013)]{zhang2013tweedie}
Yanwei Zhang.
\newblock Likelihood-based and bayesian methods for tweedie compound poisson
  linear mixed models.
\newblock \emph{Statistics and Computing}, 23:\penalty0 743--757, 2013.

\end{thebibliography}

\appendix

\section{Derivation of $\widehat{C}_n(\bmrobust{B})$}
\label{sand:app:Chat_s}
We give a series of propositions to derive $\widehat{C}_n(\matr{B})$. The
proofs all rely implicitly on the gradient and Hessian matrix of the ELBO for
variational and model parameters, given in \Cref{sand:app:derivatives-elbo}.

We recall that for a vector $x$, $\matr D_x$  denotes a diagonal matrix with diagonal $x$.  Moreover, we denote:
\begin{align*}
      \matr{\Lambda_i} & = \left( \matr{I}_p + \matr{D}_{\matr{s}_i}^2
                   (\matr{D}_{\matr{\widetilde{a}}_i} +
               \matr{D}_{\matr{\widetilde{a}_i}}\matr{D}^{2}_{\matr{s}_i} +
           \DOmeg)\right)^{-1}
           \matr{D}_{\matr{\widetilde{a}_i}}\matr{D}^{2}_{\matr{s}_i}\\
  \matr{C}_i & = \left(\matr{I}_p +
  \matr{D}_{\matr{\widetilde{a}}_i}^{-\frac12}\matr{\Omega}\matr{D}^{-\frac12}_{\matr{\widetilde{a}}_i}
- \matr{D}_{\matr{s}_i}\matr{\Lambda}_i \matr{D}_{\matr{s}_i}\right)^{-1}.
  \end{align*}
The second matrix is well defined as straightforward computations shows that
each diagonal term of the diagonal matrix
$\matr{D}_{\matr{s}_i}\matr{\Lambda}_i \matr{D}_{\matr{s}_i}$ is stricly lower
than $1$ so that $\matr I_p \succ \matr{D}_{\matr{s}_i}\matr{\Lambda}_i \matr{D}_{\matr{s}_i}$ and
\begin{equation*}\matr I_p +
  \matr{D}_{\matr{\widetilde{a}}_i}^{-\frac12}\matr{\Omega}\matr{D}^{-\frac12}_{\matr{\widetilde{a}}_i}- \matr{D}_{\matr{s}_i}\matr{\Lambda}_i
\matr{D}_{\matr{s}_i}\succ
\matr{D}_{\matr{\widetilde{a}}_i}^{-\frac12}\matr{\Omega}\matr{D}^{-\frac12}_{\matr{\widetilde{a}}_i}
\succ 0\end{equation*}
with
the partial ordering over $\mathbb S_{++}$.

\begin{proposition} \label{sand:prop:inverse-deriv-2-psi-inv}
  The inverse of the second order derivative of $J_i$ in $\matr \psi_i$ is
  \begin{multline*}
      (\nabla_{\matr \psi_i\matr \psi_i} J_{i})^{-1}(\theta, \matr \psi_i) = \\
  -\begin{pmatrix}
  \matr{D}_{\matr{\widetilde{a}}_i}^{-\frac12} &  \\
  & \matr{D}_{\matr{\widetilde{a}}_i}^{-\frac12}
  \end{pmatrix}
  \begin{pmatrix}
  \matr{C}_i & -\matr{C}_i \matr{D}_{\matr{s}_i} \matr{\Lambda}_i \\
  -\matr{\Lambda}_i \matr{D}_{\matr{s}_i} \matr{C}_i & \matr{\Lambda}_i + \matr{\Lambda}_i \matr{D}_{\matr{s}_i} \matr{C}_i \matr{D}_{\matr{s}_i} \matr{\Lambda}_i
  \end{pmatrix}
  \begin{pmatrix}
  \matr{D}_{\matr{\widetilde{a}}_i}^{-\frac12} &  \\
  & \matr{D}_{\matr{\widetilde{a}}_i}^{-\frac12}
  \end{pmatrix}.
\end{multline*}
\end{proposition}

\begin{proof} From the second order derivative of the one data ELBO
    $J_i$ in \Cref{sand:eq:derivatives-elbo-var}, we get the Hessian matrix of
    $J_i$ in $\matr \psi_i$, that is
 \begin{align*}
  (\nabla_{\matr \psi_i\matr \psi_i} J)(\theta,\matr  \psi_i) & = -
  \begin{pmatrix}
  \matr{D}_{\widetilde{\!\matr{a}}_i} + \matr{\Omega} & \matr{D}_{\widetilde{\!\matr{a}}_i}\matr{D}_{\matr{s}_i} \\
  \matr{D}_{\widetilde{\!\matr{a}}_i}\matr{D}_{\matr{s}_i} & \matr{D}_{\widetilde{\! \matr{a}}_i}(\matr{I}_p + \matr{D}^2_{\matr{s}_i}) + \matr{D}^{-2}_{\matr{s}_i}+\DOmeg
  \end{pmatrix} \\
  & = -
  \begin{pmatrix}
  \matr{D}_{\widetilde{\!\matr{a}}_i}^{1/2} &  \\
  & \matr{D}_{\widetilde{\!\matr{a}}_i}^{1/2}
  \end{pmatrix}
  \begin{pmatrix}
  \matr{I}_p + \matr{D}_{\widetilde{\!\matr{a}}_i}^{-\frac12}\matr{\Omega}\matr{D}^{-\frac12}_{\widetilde{\!\matr{a}}_i} & \matr{D}_{\matr{s}_i} \\
  \matr{D}_{\matr{s}_i} & \matr{H}_i
  \end{pmatrix}
  \begin{pmatrix}
  \matr{D}_{\widetilde{\!\matr{a}}_i}^{1/2} &  \\
  & \matr{D}_{\widetilde{\!\matr{a}}_i}^{1/2}
  \end{pmatrix},
 \end{align*}
 with $\matr{H}_i$ the $p\times p$ diagonal matrix such that
 \begin{equation*}
   \matr{H}_i = \matr{I}_p + \matr{D}^{2}_{\matr{s}_i} + \matr{D}_{\widetilde{\!\matr{a}}_i}^{-1}\matr{D}^{-2}_{\matr{s}_i} + \matr{D}_{\widetilde{\!\matr{a}}_i}^{-1}\DOmeg,
 \end{equation*}
 with
 \begin{align*} \matr{H}_i^{-1} & = \left( \matr{I}_p + \matr{D}^{2}_{\matr{s}_i} + \matr{D}_{\matr{\widetilde{a}}_i}^{-1}\matr{D}^{-2}_{\matr{s}_i} + \matr{D}_{\widetilde{\matr{a}}_i}^{-1}\DOmeg \right)^{-1}\\
                   & = \Diag\left( \frac{\widetilde{\matr{a}}_i \odot
                   \matr{s}_i^2}{\matr{1}_p + \widetilde{\matr{a}}_i \odot
           \matr{s}_i^2 + \matr{\widetilde{a}}_i \odot \matr{s}_i^4 +
   \diag(\matr{\Omega}) \odot \matr{s}_i^2} \right) \\
                   & = \left( \matr{I}_p + \matr{D}_{\matr{s}_i}^2
                   (\matr{D}_{\matr{\widetilde{a}}_i} +
               \matr{D}_{\matr{\widetilde{a}_i}}\matr{D}^{2}_{\matr{s}_i} +
           \DOmeg)\right)^{-1}
           \matr{D}_{\matr{\widetilde{a}_i}}\matr{D}^{2}_{\matr{s}_i}\\
                   & = \matr{\Lambda}_i
 \end{align*}
 Tedious but straightforward computations stemming from blockwise diagonal matrices lead to the inverse.
\end{proof}

\begin{lemma} \label{sand:lem:kro-diag}
    Let $\matr{A}$ and $\matr{D}$ be square matrices of size $p$ and $\matr{x}$ a column-vector of length $m$, we have
\begin{align*}
    (\matr{D} \kro \matr{x}) \matr{A}& = \left(\matr{DA}\right) \kro \matr{x}\\
 (\matr{x}\kro \matr{D}) \matr{A}& = \matr{x} \kro \left(\matr{DA}\right)
\end{align*}
\end{lemma}
\begin{proof}
    This is a direct consequence of the matrix equality $(B\kro C) (E\kro F) = (BE)\kro(CF)$.
\end{proof}

\begin{proposition} \label{sand:prop:inverse-deriv-2-theta-psi-theta}
    With the previous notations,
 \begin{align*}
 \left[ \nabla_{\text{vec}(\matr{B})\matr \psi_i} J_{i} \left( \nabla_{\matr
 \psi_i\matr \psi_i} J_{i} \right)^{-1} \nabla_{\matr
\psi_i\text{vec}(\matr{B})} J_{i} \right](\theta, \matr \psi_i) & =  -
(\matr{D}_{\widetilde{\!\matr{a}}_i}^{1/2} \matr{E}_i
\matr{D}_{\widetilde{\!\matr{a}}_i}^{1/2}) \kro \matr{x}_i \matr{x}_i^\top
 \end{align*}
 where
 $\matr{E}_i = \bG_i +
 \left(\matr{I}_p -
 \bG_i\right)  \matr{C}_i
 \left(\matr{I}_p -
 \bG_i\right)$ and $\bG_i=\matr{D}_{\matr{s}_i}\matr{\Lambda}_i\matr{D}_{\matr{s}_i}$.
\end{proposition}

\begin{proof}
Using \Cref{sand:lem:kro-diag}, we get
 \[
\begin{bmatrix}
\matr{D}_{\widetilde{\!\matr{a}}_i} \kro \matr{x}_i, &
\matr{D}_{\widetilde{\!\matr{a}}_i\odot \matr{s}_i} \kro \matr{x}_i
\end{bmatrix}
\begin{bmatrix}
\matr{D}_{\widetilde{\!\matr{a}}_i}^{-1/2} & \\
& \matr{D}_{\widetilde{\!\matr{a}}_i}^{-1/2}
\end{bmatrix} =
\begin{bmatrix}
\matr{D}_{\widetilde{\!\matr{a}}_i}^{1/2} & \matr{D}_{\widetilde{\!\matr{a}}_i}^{1/2}\matr{D}_{\!\matr{s}_i}
\end{bmatrix} \kro \matr{x}_i.
\]
Then, using \Cref{sand:prop:inverse-deriv-2-psi-inv}, the gradient formula of the cross derivative in \Cref{sand:prop:inverse-deriv-2-psi-theta}, another application of
lemma~\ref{sand:lem:kro-diag} and the mixed-product property of the kronecker
product, one has
\begin{align*}
 & \left( \begin{bmatrix}
\matr{D}_{\widetilde{\!\matr{a}}_i}^{1/2} & \matr{D}_{\widetilde{\!\matr{a}}_i}^{1/2}\matr{D}_{\!\matr{s}_i}
\end{bmatrix} \kro \matr{x}_i
\right)
\begin{bmatrix}
\matr{C}_i & -\matr{C}_i\matr{D}_{\matr{s}_i}\matr{\Lambda}_i \\
-\matr{\Lambda}_i \matr{D}_{\matr{s}_i}\matr{C}_i & \matr{\Lambda}_i + \matr{\Lambda}_i \matr{D}_{\matr{s}_i}\matr{C}_i \matr{D}_{\matr{s}_i}\matr{\Lambda}_i
\end{bmatrix}
\left( \begin{bmatrix}
\matr{D}_{\widetilde{\!\matr{a}}_i}^{1/2} \\ \matr{D}_{\widetilde{\!\matr{a}}_i}^{1/2}\matr{D}_{\!\matr{s}_i}
\end{bmatrix} \kro \matr{x}_i^\top
\right)
\\
 & =
\begin{bmatrix}
\matr{D}_{\widetilde{\!\matr{a}}_i}^{1/2} & \matr{D}_{\widetilde{\!\matr{a}}_i}^{1/2}\matr{D}_{\!\matr{s}_i}
\end{bmatrix}
\begin{bmatrix}
\matr{C}_i & -\matr{C}_i\matr{D}_{\matr{s}_i}\matr{\Lambda}_i \\
-\matr{\Lambda}_i \matr{D}_{\matr{s}_i}\matr{C}_i & \matr{\Lambda}_i + \matr{\Lambda}_i \matr{D}_{\matr{s}_i}\matr{C}_i \matr{D}_{\matr{s}_i}\matr{\Lambda}_i
\end{bmatrix}
\begin{bmatrix}
\matr{D}_{\widetilde{\!\matr{a}}_i}^{1/2} \\ \matr{D}_{\widetilde{\!\matr{a}}_i}^{1/2}\matr{D}_{\!\matr{s}_i}
\end{bmatrix}
\kro \left(  \matr{x}_i \matr{x}_i^\top \right)\\
 & = \matr{D}_{\widetilde{\!\matr{a}}_i}^{1/2} \left(
  \bG_i - \bG_i\matr{C}_i - \matr{C}_i\bG_i  + \matr{C}_i + \bG_i\matr{C}_i\bG_i
\right)  \matr{D}_{\widetilde{\!\matr{a}}_i}^{1/2}\\
 & =
\matr{D}_{\widetilde{\!\matr{a}}_i}^{1/2} \left(
  \bG_i + \left(\matr{I}_p - \bG_i\right)  \matr{C}_i \left(\matr{I}_p - \bG_i\right)\right)  \matr{D}_{\widetilde{\!\matr{a}}_i}^{1/2}.
\end{align*}

\end{proof}

\begin{proposition} \label{sand:prop:inverse-deriv-2-theta-theta}
 With the previous notations,
 \begin{multline*}
     \left[ \nabla_{\text{vec}(\matr{B})\text{vec}(\matr{B})} J_{i} - \nabla_{\text{vec}(\matr{B})\matr{\psi}_i} J_{i} \left( \nabla_{\matr{\psi}_i\matr \psi_i} J_{i} \right)^{-1} \nabla_{\matr \psi_i\text{vec}(\matr{B})} J_{i} \right](\theta, \matr \psi_i) \\
   = - \left(\matr{\Sigma} + \matr{D}_{\widetilde{\matr{a}}_i}^{-1} + \matr{D}_{\matr{s}_i}^4 \left( \matr{I}_p + \matr{D}_{\matr{s}_i}^2 (\matr{D}_{\widetilde{\matr{a}}_i} + \DOmeg)\right)^{-1} \right)^{-1} \kro \left(  \matr{x}_i \matr{x}_i^\top \right)
 \end{multline*}
\end{proposition}

\begin{proof} By Prop.~\ref{sand:prop:inverse-deriv-2-theta-psi-theta} and Equation~\ref{sand:prop:inverse-deriv-2-theta}, the expression we have to expand and simplify is
  \begin{equation*}
    \matr{D}_{\widetilde{\!\matr{a}}_i}^{1/2}(\matr{I}_p - \matr{E}_i)\matr{D}_{\widetilde{\!\matr{a}}_i}^{1/2}.
  \end{equation*}
Denote $\widetilde{\matr{\Lambda}} = \matr{I}_p - \matr{D}_{\matr{s}_i}
\matr{\Lambda}_i \matr{D}_{\matr{s}_i}$ and $\widetilde{\matr{\Omega}} =
\matr{D}_{\widetilde{\!\matr{a}}_i}^{-1/2}\matr{\Omega}\matr{D}_{\widetilde{\!\matr{a}}_i}^{-1/2}$
and $\matr{C} = (\widetilde{\matr{\Omega}} +
\widetilde{\matr{\Lambda}})^{-1}$. We have
  \begin{align*}
    \matr{I}_p - \matr{E}_i  & = \matr{I}_p - \matr{D}_{\matr{s}_i} \matr{\Lambda}_i \matr{D}_{\matr{s}_i} - (\matr{I}_p -\matr{D}_{\matr{s}_i} \matr{\Lambda}_i \matr{D}_{\matr{s}_i}) \matr{C} (\matr{I}_p - \matr{D}_{\matr{s}_i} \matr{\Lambda} \matr{D}_{\matr{s}_i}) \\
                             & = (\matr{I}_p - \matr{D}_{\matr{s}_i} \matr{\Lambda} \matr{D}_{\matr{s}_i}) ( (\matr{I}_p - \matr{D}_{\matr{s}_i} \matr{\Lambda} \matr{D}_{\matr{s}_i})^{-1} - \matr{C}_i) (\matr{I}_p - \matr{D}_{\matr{s}_i} \matr{\Lambda} \matr{D}_{\matr{s}_i}) \\
                             & = \widetilde{\matr{\Lambda}} ( \widetilde{\matr{\Lambda}}^{-1} - (\widetilde{\matr{\Omega}} + \widetilde{\matr{\Lambda}})^{-1}) \widetilde{\matr{\Lambda}} \\
                             & = \widetilde{\matr{\Lambda}} - \widetilde{\matr{\Lambda}} (\widetilde{\matr{\Omega}} + \widetilde{\matr{\Lambda}})^{-1} \widetilde{\matr{\Lambda}}  \\
    & = \left(\widetilde{\matr{\Omega}}^{-1} + \widetilde{\matr{\Lambda}}^{-1}\right)^{-1},
  \end{align*}
  where the last line stands by the Woodbury identity. By the  Expression of $\matr{\Lambda}_i$, we have
  \begin{equation*}
    \widetilde{\matr{\Lambda}} =  \matr{I}_p - \matr{D}_{\matr{s}_i}
    \matr{\Lambda}_i \matr{D}_{\matr{s}_i} = \left( \matr{I}_p +
    \matr{D}_{\matr{s}_i}^2 (\matr{D}_{\widetilde{\!\matr{a}}_i} +
\matr{D}_{\widetilde{\!\matr{a}}_i}\matr{D}^{2}_{\matr{s}_i} +
\DOmeg)\right)^{-1} \left( \matr{I}_p + \matr{D}_{\matr{s}_i}^2
(\matr{D}_{\widetilde{\!\matr{a}}_i} + \DOmeg)\right) \, \\
  \end{equation*}
  and
  \begin{align*}
    \widetilde{\matr{\Lambda}}^{-1} & =  \left( \matr{I}_p + \matr{D}_{\matr{s}_i}^2 (\matr{D}_{\widetilde{\!\matr{a}}_i} + \DOmeg)\right)^{-1} \left( \matr{I}_p + \matr{D}_{\matr{s}_i}^2 (\matr{D}_{\widetilde{\!\matr{a}}_i} + \matr{D}_{\widetilde{\!\matr{a}}_i}\matr{D}^{2}_{\matr{s}_i} + \DOmeg)\right) \\
                                & = \matr{I}_p + \matr{D}_{\widetilde{\!\matr{a}}_i}\matr{D}_{\matr{s}_i}^4 \left( \matr{I}_p + \matr{D}_{\matr{s}_i}^2 (\matr{D}_{\widetilde{\!\matr{a}}_i} + \DOmeg)\right)^{-1}.
  \end{align*}
Finally, we are left with
  \begin{align*}
    \matr{D}_{\widetilde{\!\matr{a}}_i}^{1/2}(\matr{I}_p - \matr{E}_i)\matr{D}_{\widetilde{\!\matr{a}}_i}^{1/2} & = \matr{D}_{\widetilde{\!\matr{a}}_i}^{-1/2}\left(\widetilde{\matr{\Omega}}^{-1} + \widetilde{\matr{\Lambda}}^{-1}\right)^{-1}\matr{D}_{\widetilde{\!\matr{a}}_i}^{-1/2}\\
    & = \matr{D}_{\widetilde{\!\matr{a}}_i}^{1/2}\left( \matr{D}_{\widetilde{\!\matr{a}}_i}^{1/2} \matr{\Omega}^{-1}\matr{D}_{\widetilde{\!\matr{a}}_i}^{1/2} + \widetilde{\matr{\Lambda}}^{-1}\right)^{-1}\matr{D}_{\widetilde{\!\matr{a}}_i}^{1/2}\\
    & = \left(  \matr{\Omega}^{-1} + \matr{D}_{\widetilde{\!\matr{a}}_i}^{-1/2}\widetilde{\matr{\Lambda}}^{-1}\matr{D}_{\widetilde{\!\matr{a}}_i}^{-1/2}\right)^{-1}\\
    & = \left(\matr{\Sigma} + \matr{D}_{\widetilde{\!\matr{a}}_i}^{-1} + \matr{D}_{\matr{s}_i}^4 \left( \matr{I}_p + \matr{D}_{\matr{s}_i}^2 (\matr{D}_{\widetilde{\!\matr{a}}_i} + \DOmeg)\right)^{-1} \right)^{-1}.
\end{align*}

\end{proof}

\section{Gradient and Hessian Matrix of the ELBO} \label{sand:subsec:gradient}
\label{sand:app:derivatives-elbo}
\paragraph{Model parameters}
The first and second-order derivative (gradient and Hessian matrix) of the
ELBO w.r.t the vector of the model parameters
$\theta = (\matr{B}, \matr{\Omega})$ are given by
\begin{align*}
  \begin{split}
      \nabla_{\matr{B}} \barJ(\theta, \matr \psi)  & = \matr{X}^\top (\matr{Y}  - \matr{\widetilde{A}}), \\
    \nabla_{\matr{\Omega}} \barJ(\theta, \matr \psi) & = \frac{n}{2} \left[  \matr{\Omega}^{-1} - \frac{\matr{M}^\top\matr{M} + \bar{\matr{S}}^2}{n} \right], \\
    \nabla_{\matr{B}\matr{B}} \barJ(\theta, \matr \psi)  & = -(\matr{I}_p \kro \matr{X}^\top)\matr{D}_{\opvec(\matr{\widetilde{A}})}(\matr{I}_p \kro \matr{X}), \\[1.5ex]
    \nabla_{\matr{\Omega}\matr{B}} \barJ(\theta, \matr \psi)  & = \matr{0}_{mp, p^2}, \\[1.5ex]
    \nabla_{\matr{\Omega}\matr{\Omega}} \barJ(\theta, \matr \psi)  & = -\frac{n}{2}\matr{\Omega}^{-1} \kro
\matr{\Omega}^{-1}.
  \end{split}
\end{align*}
The Hessian matrix is block-diagonal in $(\matr{B}, \matr{\Omega})$ and given by
\begin{equation*}
H(\theta, \matr \psi) = \begin{pmatrix}
    -(\matr{I}_p \kro \matr{X}^\top)\matr{D}_{\opvec(\widetilde{A})}(\matr{I}_p \kro \matr{X}) &  \matr{0}_{mp, p^2} \\
  \matr{0}_{mp, p^2}  & -\frac{n}{2}\matr{\Omega}^{-1} \kro \matr{\Omega}^{-1}
\end{pmatrix}.
\end{equation*}

\paragraph{Variational parameters}
The first and second order derivative (gradient and Hessian matrix) of the complete
likelihood w.r.t the vector of the variational parameters $\matr{\psi} = (\matr \psi_1, \dots, \matr \psi_n)$ are given in the following. For $i \in \{1,\dots,n\}$,
\begin{alignat}{10}
  \nabla_{\matr{m}_i} J_i(\ta, \matr\psi_i) &\hspace*{2\arraycolsep}&=&\hspace*{2\arraycolsep} \matr{Y}_i - \widetilde{\!\matr{a}}_i - \matr{\Omega}\matr{m}_i \nonumber, \\[1.5ex]
    \nabla_{\matr{s}_i} J_i(\ta, \matr\psi_i)  &\hspace*{2\arraycolsep}&=&\hspace*{2\arraycolsep}- \matr{s}_i \matr{D}_{\widetilde{\!\matr{a}}_i} + 1/\matr{s}_i - \matr{s}_i\DOmeg,\nonumber \\[1.5ex]
    \nabla_{\matr{m}_i\matr{m}_i} J_i(\ta, \matr\psi_i) &\hspace*{2\arraycolsep}&=&\hspace*{2\arraycolsep}- \matr{D}_{\widetilde{\matr{a}}_i} - \matr{\Omega}, \label{sand:eq:derivatives-elbo-var} \\[1.5ex]
    \nabla_{\matr{m}_i\matr{s}_i} J_i(\ta, \matr\psi_i) &\hspace*{2\arraycolsep}&=&\hspace*{2\arraycolsep}- \matr{D}_{\widetilde{\matr{a}}_i}\matr{D}_{\matr{s}_i} = -\matr{D}_{\widetilde{\matr{a}_i} \odot \matr{s}_i}, \nonumber \\[1.5ex]
    \nabla_{\matr{s}_i\matr{s}_i} J_i(\ta, \matr\psi_i) &\hspace*{2\arraycolsep}&=&\hspace*{2\arraycolsep}-\matr{D}_{\widetilde{\! \matr{a}_i}}(\matr{I}_p + \matr{D}^2_{\matr{s}_i}) - \matr{D}^{-2}_{\matr{s}_i}-\DOmeg. \nonumber
    \intertext{In matrix form, we have}
 \nabla_{\matr{M}} \barJ(\ta, \matr\psi) &\hspace*{2\arraycolsep}&=&\hspace*{2\arraycolsep} \matr{Y} - \widetilde{\!\matr{A}} - \matr{M}\matr{\Omega},\nonumber \\[1.5ex]
 \nabla_{\matr{S}} \barJ(\ta, \matr\psi) &\hspace*{2\arraycolsep}&=&\hspace*{2\arraycolsep} -\matr{S} \odot \widetilde{\!\matr{A}} + 1/\matr{S} - \matr{S}\DOmeg, \nonumber\\[1.5ex]
\nabla_{\matr{M}\matr{M}} \barJ(\ta, \matr\psi) &\hspace*{2\arraycolsep}&=&\hspace*{2\arraycolsep} - \matr{D}_{\opvec(\widetilde{\!\matr{A}})} - \matr{I}_n \kro \matr{\Omega}, \nonumber \\[1.5ex]
 \nabla_{\matr{M}\matr{S}} \barJ(\ta, \matr\psi) &\hspace*{2\arraycolsep}&=&\hspace*{2\arraycolsep} - \matr{D}_{\opvec(\widetilde{\!\matr{A}})}\matr{D}_{\opvec(\matr{S})},\nonumber \\[1.5ex]
    \nabla_{\matr{S}\matr{S}} \barJ(\ta, \matr\psi) &\hspace*{2\arraycolsep}&=&\hspace*{2\arraycolsep} -\matr{D}_{\opvec(\widetilde{\!A})}(\matr{I}_{np} + \matr{D}^2_{\opvec(\matr{S})})  - \matr{D}^{-2}_{\opvec(\matr{S})}-\matr{I}_n\kro \DOmeg. \nonumber
\end{alignat}

\paragraph{Additional results for vectorized forms of derivatives}
We report here some closed-form expressions of the first and second derivative
of a single data variational ELBO $J_{i}$ for $\matr{B}$.
\begin{align}
  \label{sand:prop:inverse-deriv-1-Theta}
  \nabla_{\text{vec}(\matr{B})} J_i(\theta, \matr \psi_i) & = (\matr{Y}_i - \matr{\widetilde{a}}_i) \kro \matr{x}_i \in \mathcal{\mathbb R}^{mp \times 1}, \\
    \label{sand:prop:inverse-deriv-2-theta}
  \nabla_{\text{vec}(\matr{B})\text{vec}(\matr{B})} J_i(\theta, \matr \psi_i)&  = - \matr{D}_{\widetilde{\!\matr{a}}_i} \kro \left(  \matr{x}_i \matr{x}_i^\top \right) \in \mathbb R^{pm \times pm}.
\end{align}
Finally, we give the cross derivative between $\text{vec}(\matr{B})$ and the variational parameters of a single observation:
\begin{equation}
  \label{sand:prop:inverse-deriv-2-psi-theta}
  \nabla_{\opvec(\matr{B})\matr \psi_i} J_i(\theta, \matr \psi_i) = - \begin{bmatrix}
    {\underbrace{\matr{D}_{\widetilde{\!\matr{a}}_i} \kro \matr{x}_i}_{\in \mathbb R^{pm \times p}}} &
    {\underbrace{ \matr{D}_{\widetilde{\!\matr{a}_i}\odot \matr{s}_i} \kro \matr{x}_i}_{\in \mathbb R^{pm \times p}}}
  \end{bmatrix}.
\end{equation}

\end{document}